\documentclass[%
twocolumn,
nofootinbib,
amsmath,amssymb,
aps,
pre
]{revtex4-2}

\usepackage{graphicx,xcolor}
\usepackage{dcolumn}
\usepackage{bm}
\usepackage{amsmath,amsfonts,amssymb}

\begin{document}

\preprint{APS/123-QED}

\title{{\it Amoeba} Monte Carlo algorithms for random trees with controlled branching activity: efficient trial move generation and universal dynamics}

\author{Pieter H. W. van der Hoek}
 \email{pvanderh@sissa.it}
\affiliation{%
 SISSA - Scuola Internazionale Superiore di Studi Avanzati, Via Bonomea 265, 34136 Trieste, Italy
}%

\author{Ralf Everaers}
    \email{ralf.everaers@ens-lyon.fr}
\affiliation{
ENS de Lyon, CNRS, Laboratoire de Physique (LPENSL UMR5672) et Centre Blaise Pascal, 69342 Lyon cedex 07, France
}%

\author{Angelo Rosa}
\email{anrosa@sissa.it}
\affiliation{%
 SISSA - Scuola Internazionale Superiore di Studi Avanzati, Via Bonomea 265, 34136 Trieste, Italy
}%

\date{\today}

%
\begin{abstract}
The reptation Monte Carlo algorithm is a simple, physically motivated and efficient method for equilibrating semi-dilute solutions of linear polymers.
Here we propose two simple generalizations for the analogue {\it Amoeba} algorithm for randomly branching chains, which allow to efficiently deal with random trees with controlled branching activity.
We analyse the rich relaxation dynamics of Amoeba algorithms and demonstrate the existence of an unexpected scaling regime for the tree relaxation.
In particular, our results suggests that the equilibration time for Amoeba algorithms scales in general like $N^2 \langle n_{\rm lin}\rangle^\Delta$, where $N$ denotes the number of tree nodes, $\langle n_{\rm lin}\rangle$ the mean number of linear segments the trees are composed of and $\Delta \simeq 0.4$.
\end{abstract}
%

\maketitle


\section{Introduction}\label{sec:Intro}
Branched polymers or ``trees'' without cyclic units~\cite{RubinsteinColby,FisherPRL1978,KurtzeFisherPRB1979,IsaacsonLubensky,DuarteRuskin1981,ParisiSourlasPRL1981,BovierFroelichGlaus1984,Burchard1999}  have proven to be remarkably successful in the past decades as physical models for a wide range of systems. 
Particularly interesting is the use of trees with {\it annealed connectivity}~\cite{GutinGrosberg93} to describe the packing of long unknotted and non-concatenated ring polymers in melt conditions~\cite{KhokhlovNechaev85,RubinsteinPRL1986,RubinsteinPRL1994,GrosbergSoftMatter2014,RosaEveraersPRL2014,MichielettoSoftMatter2016,RosaEveraers2019,SchramSM2019,SmrekRosa2019,Ghobadpour2021}, the classification of long viral RNA molecules~\cite{RNA_ensemble,RNA_virus,PrueferAlgoRNA,Vaupotic2023} and the double-folded structures of bacterial DNA~\cite{MarkoSiggia1994,MarkoSiggiaSuperCoiledDNA1995,odijk_osmotic_1998,Cunha2001,Junier2023} and chromosomal DNA during interphase~\cite{grosbergEPL1993,RosaPLOS2008,Vettorel2009,MirnyReview2011,halverson2014melt,WeisshaarBJ2011,EveraersGrosbergRubinsteinRosa,JOST2018}.
In these contexts, trees are employed as coarse-grained descriptions of the structures of interest.
These minimal representations of polymers and biopolymers have a small parameter space and allow to systematically explore the rich large-scale behavior. 
Due to the small number of available exact results~\cite{ZimmStockmayer49,DeGennes1968,DaoudJoanny1981,GrosbergNechaev2015,LubenskyIsaacson1979,ParisiSourlasPRL1981} most problems in the field call for numerical studies~\cite{Redner1979,SeitzKlein1981,Caracciolo1985,MadrasJPhysA1992,GrassbergerJPhysA2005} as a complement to approximate theoretical treatments~\cite{deGennes76,KurtzeFisherPRB1979,DeGennesThetaCollapse1975,LeGuillouZinnJustin,Family1980,deAlcantaraMcKane1980,DerridaDeseze1982,NienhuisSAW1982,DharPRL1983,DuplantierJPhysA1986,DuplantierSaleurPRL1987,AdlerPRB1988,JanssenStenullPRE2011,EveraersGrosbergRubinsteinRosa,Marcato2023}.

While there exist other Monte Carlo (MC) algorithms~\cite{Caracciolo1985,MadrasJPhysA1992,GrassbergerJPhysA2005} for generating isolated trees, the {\it Amoeba} algorithm~\cite{SeitzKlein1981} is a simple and natural choice for simulating dense systems~\cite{Rosa2016b}.
It derives from the {\it reptation} (or {\it slithering-snake} algorithm~\cite{Wall}) for semi-dilute solutions of linear polymers, where trial moves consists in first removing a chain segment at a randomly chosen chain end and then adding it in a randomly chosen orientation to the same or the other chain end.
For a chain of length $N$ the memory of the original chain conformation is lost after ${\mathcal O}(N^2)$ MC steps. 
The Amoeba algorithm~\cite{SeitzKlein1981} generalizes this idea to branched polymers. 
Again, the first step of constructing a MC trial move consists in removing a randomly chosen end segment (or ``leaf'') from the current tree.
In the second step, the segment is added to a randomly chosen tree node.
Since the number of end and branch-points changes as a function of the functionality of the involved tree nodes, some trivial bookkeeping is required for the acceptance probabilities to observe the detailed balance condition.
Otherwise the Amoeba algorithm is expected~\cite{MadrasJPhysA1992} to be as efficient for random trees as the reptation algorithm for linear chains and to decorrelate tree conformations in ${\mathcal O}(N^2)$ MC steps.

In the present paper, we explore how to efficiently treat the case where the average number of linear segments between branch-points is controlled by a branching activity or a corresponding chemical potential $\mu$~\cite{Rosa2016a,Rosa2016b,Rosa2016c} favoring or suppressing the presence of branch-points. 
On first sight, one might expect this to be of little consequence for the efficiency of the Amoeba algorithm.
After all, there are even more end nodes to choose from in the limit of strong branching, while one might hope to recover the equally good performance of the reptation algorithm when dealing with essentially linear chains in the opposite limit of no branching. 
However, closer inspection reveals a strong decrease of the acceptance probability for the classical Amoeba moves.
For example, when branching is suppressed, most attempts to attach the severed segment to one of the inner nodes of the truncated linear chain are rejected.
It turns out that the situation is even worse for strongly branching trees with an even number of nodes, which need to pass through an exponentially suppressed {\it excited state} containing a linear chain segment.
Here, we show how to modify the Amoeba algorithm to render its efficiency independent of $\mu$ and present a detailed analysis of the relaxation dynamics.

The paper is organized as follows:
In Sec.~\ref{sec:ModelMethods}, we introduce the lattice tree model together with some structural observables and corresponding theoretical results.
In Sec.~\ref{sec:Methods}, we provide the numerical recipes for the MC algorithms under discussion, namely the {\it original} (single-leaf) Amoeba algorithm (Sec.~\ref{sec:ConventionalAmoeba}), the {\it semi-kinetic} (single-leaf) Amoeba algorithm (Sec.~\ref{sec:SemiKinAmoeba}) and the (semi-kinetic) {\it double-leaf} Amoeba algorithm (Sec.~\ref{sec:IntroduceDL}).
We furthermore define the equilibration time and rederive the expected ${\mathcal O}(N^2)$ scaling (Sec.~\ref{sec:Equilibration}).
In Sec.~\ref{sec:Results}, we validate the algorithms and their implementation by comparing obtained results between different algorithms and to theory.
In particular, we report how the different algorithms perform for different branching probabilities as well as different chain lengths.
The detailed discussion of our results in Sec.~\ref{sec:Discussion} begins (Sec.~\ref{sec:RelativePerformance}) with a presentation of the relative performance of the different algorithms and continues (Sec.~\ref{sec:MovesEfficiency}) by linking the performance to the efficient generation of trial moves.
Then we show (Sec.~\ref{sec:Universal_dynamics}) that chain dynamics obeys universal behavior that can be rationalized within a scaling description identifying relevant length and time scales.
Further variations and possible generalizations of our methods are discussed in Sec.~\ref{sec:Generalizations}. 
Finally we conclude, with an outlook on future research, in Sec.~\ref{sec:Conclusion}.

\section{Theoretical background}\label{sec:ModelMethods}
In Section~\ref{sec:Model} we briefly present the employed lattice tree model from Refs.~\cite{Rosa2016a,Rosa2016b,Rosa2016c}, followed by
the definition of the main observables (Sec.~\ref{sec:Observables}) for measuring tree size and connectivity, and a brief summary of a number of useful theoretical results (Sec.~\ref{sec:DaoudJoanny}).

\subsection{The model: definitions, branching Hamiltonian}\label{sec:Model}
We model branching polymers in $d=3$ spatial dimensions as {\it trees} on the {\it face-centered cubic} (FCC) lattice.
A tree ${\mathcal T} \equiv {\mathcal T} (\mathcal{G}, \Gamma)$, consisting of $N$ connected nodes, is described in terms of its connectivity ${\mathcal G}$ and the spatial positions of the nodes $\Gamma = \{ {\vec r}_1, \dots, {\vec r}_N \}$ on the lattice.
The nodes are simply connected by $N-1$ bonds, {\it with no internal loops}.
Here we consider completely flexible bonds, {\it i.e.} the Kuhn length~\cite{RubinsteinColby} of the polymers $\ell_{K} = a$ where $a$ is the unit lattice spacing.
In the present work we consider {\it ideal} trees with no excluded-volume, namely distinct tree nodes can occupy the same lattice sites. 

Then, we classify nodes based on their {\it functionality}, that is defined as the number of nodes directly connected to that node.
Similarly to~\cite{Rosa2016a,Rosa2016b,Rosa2016c} and without loss of generality, the functionality of the nodes is restricted here to $f\leq 3$ (for generalization to $f>3$, see also the discussion in Sec.~\ref{sec:NodesUnrestrictedFuncts}).
Our trees then feature three classes of nodes: an end-point or leaf has $f=1$, a linear segment has $f=2$, and a branch-point has $f=3$.

For any given number of $1$, $2$, and $3$-functional nodes ($n_1$, $n_2$ and $n_3$, respectively) with, obviously, $N = n_1 + n_2 + n_3$, we have that the explicit non-cyclic topology of trees allows to express $n_1$ and $n_2$ as a function of $N$ and $n_3$:
\begin{eqnarray}
n_1 & = & n_3 + 2 \, , \label{n3n1} \\
n_2 & = & N - 2n_3 - 2 \, . \label{n3n2}
\end{eqnarray}
A tree can thus be classified by $N$ and its number of branch-points $n_3$.
In order to control the branching activity in a tree, we introduce the Hamiltonian~\cite{Rosa2016a,Rosa2016b,Rosa2016c}:
\begin{equation}\label{eq:Hamiltonian}
{\mathcal H}({\mathcal T}) = -\mu \, n_3 ({\mathcal T}) \equiv -\mu \, n_3 ({\mathcal G}) \, ,
\end{equation}
where $\mu$ is the external ``chemical'' potential that, depending on its sign and magnitude, is used to tune the amount of branching in a tree.
In the end, as we will see, $\mu$ may either define a length scale or a weight: namely, the average length of linear sections between branch-points (or number of ``$f\!=\!2$''-nodes) for $\mu \ll 0$ and the average tree weight per ``$f\!=\!2$''-node for $\mu \gg 0$.

Notice that since the Hamiltonian~\eqref{eq:Hamiltonian} depends solely on the connectivity ${\mathcal G}$ of the tree (and no other interaction like, for instance, excluded-volume is considered in this work), so does the expectation value,
\begin{equation}\label{eq:MeanValueGenericObs}
\langle A \rangle \equiv \sum_{{\mathcal G}_i} A({\mathcal T}_i) \, w({\mathcal G}_i) \, , \, \, \, w({\mathcal G}_i) \equiv \frac{ e^{-\beta {\mathcal H({\mathcal G}_i)}} }{ \sum_{{\mathcal G}_i} e^{-\beta {\mathcal H({\mathcal G}_i)}} } \, ,
\end{equation}
of {\it any} tree observable $A = A({\mathcal T})$ where $\beta=1 / (k_BT)$ is the usual Boltzmann factor at temperature $T$.
In particular the expectation value~\eqref{eq:MeanValueGenericObs} does not depend on the spatial dimensionality $d$ nor on the specific choice of the underlying lattice, and we can  compare then (Sec.~\ref{sec:Benchmarking}) the results of this work to those for ideal trees published in~\cite{Rosa2016a,Rosa2016b,Rosa2016c} where the simple cubic lattice was used instead of the FCC one.

\subsection{Tree observables}\label{sec:Observables}
In order to test both efficiency and accuracy of the different algorithms, we introduce two fundamental observables~\cite{Rosa2016a,Rosa2016b,Rosa2016c}.
The first quantity, $\left \langle n_3 \right \rangle$, is the tree mean number of branch-points, while the second is the tree mean-square gyration radius (in units of square lattice spacing $a^2$, see Sec.~\ref{sec:Model}), 
\begin{equation}\label{eq:MeanSqGyrRadius}
\langle R_g^2 \rangle = \left\langle \frac{1}{2N^2} \sum_{i\neq j} ( {\vec r}_i - {\vec r}_j )^2 \right\rangle \, ,
\end{equation}
that quantifies polymer size.
Then, for different values of $\mu$ in the Hamiltonian~\eqref{eq:Hamiltonian}, different ensembles should be expected:
\begin{itemize}
\item
$\mu / k_BT \ll 0$.
It is energetically highly unfavourable to introduce branching in a given tree and linear chains dominate the ensemble.
\item
$\mu / k_BT \gg 0$.
Trees are {\it maximally branched}.
Due to the linear relationship between $n_1$ and $n_3$ (Eq.~\eqref{n3n1}), the maximal number of branch-points $n_{3, {\rm max}}$ can be written as:
\begin{eqnarray}
n_{3, {\rm max}} & = & (N-2)/2 \, , \, \, \, {\rm for} \, N \, {\rm even} \, , \label{n3maxNeven} \\
n_{3, {\rm max}} & = & (N-3)/2 \, , \, \, \, {\rm for} \, N \, {\rm odd} \, . \label{n3maxNodd}
\end{eqnarray}
There is thus a (small) difference between chains with an even number of monomers and chains with an odd number of monomers $N$.
In particular, a chain with an odd number of monomers has {\it always} one node with $f\!=\!2$ in the maximally branched state.
A chain with an even number of monomers has no such node.
\item
$\mu / k_BT \sim {\mathcal O}(1)$.
For random branching, entropic and energetic effects determine together which structures are dominant in a constant $\mu$-ensemble.
\end{itemize}
Polymers with higher branching activity are in general more compact objects.
In particular, given $N$, the mean-square gyration radius for linear chains obeys $\left \langle R_g^2 \right \rangle \sim N^1$~\cite{RubinsteinColby}, while for highly branched structures in the large $N$-limit we do have $\left \langle R_g^2 \right \rangle \sim N^{1/2}$~\cite{ZimmStockmayer49}.
In the next Section, we return on these exact results in the context of the theoretical work by Daoud and Joanny~\cite{DaoudJoanny1981}.

\subsection{Daoud-Joanny theory for ideal random trees}\label{sec:DaoudJoanny}
In a seminal paper~\cite{DaoudJoanny1981}, Daoud and Joanny computed the {\it exact} partition function, ${\mathcal Z}_{N_b}$, for the ensemble of ideal trees made of $N_b=N-1$ elastic flexible bonds in the continuum approximation:
\begin{equation}\label{eq:DJpartfunct}
{\mathcal Z}_{N_b} = \frac{I_1 \left( 2 \, \lambda \, N_b \right) }{\lambda \, N_b} \simeq
\left\{
\begin{array}{cc}
1 + \frac{(\lambda N_b)^2}{2} , \, & \lambda N_b \ll 1 \\
\\
\frac{e^{2 \lambda N_b}}{2 \pi^{1/2} (\lambda N_b)^{3/2}}, \, & \lambda N_b \gg 1
\end{array}
\right. \, ,
\end{equation}
where $I_1(x)$ is the first modified Bessel function of the first kind, $\lambda$ the branching fugacity (related to the activity of the trifunctional units and the chemical potential $\mu$) and ${\mathcal Z}_{N_b=0} = 1$.

All interesting properties of ideal trees can be derived by means of Eq.~\eqref{eq:DJpartfunct}.
For instance, by removing a randomly chosen bond, splits a branch of size $n_b < N_b/2$ from the remaining tree of size $N_b-n_b-1$.
Then, the mean-square gyration radius $\langle R_g^2(N_b) \rangle$ in bond units is given by the famous Kramers theorem~\cite{Kramers1946,RubinsteinColby}:
\begin{widetext}
\begin{equation}\label{eq:DaoudJoanny-Rg2}
\langle R_g^2(N_b) \rangle
= \frac{N_b}{(N_b+1)^2} \frac{ \sum_{n_b=0}^{N_b-1} (n_b+1)(N_b-n_b) {\mathcal Z}_{n_b} \, {\mathcal Z}_{N_b-1-n_b} } { \sum_{n_b=0}^{N_b-1} \, {\mathcal Z}_{n_b} \, {\mathcal Z}_{N_b-1-n_b} }
\simeq
\left\{
\begin{array}{lc}
\frac{N_b}6 \frac{N_b+2}{N_b+1} \, , & \lambda N_b \ll 1 \\
\\
\frac{\sqrt{\pi}}4 \sqrt{\frac{N_b}{\lambda}} \, , & \lambda N_b \gg 1
\end{array}
\right. \, .
\end{equation}
\end{widetext}
Similarly, the mean number of branching monomers is given by the expression:
\begin{equation}\label{eq:DaoudJoanny-n3}
\langle n_3(N_b) \rangle
= \frac{N_b}2 \frac{\partial{\log( \mathcal Z}_{N_b} )}{\partial N_b}
\simeq
\left\{
\begin{array}{lc}
\frac{(\lambda N_b)^2}2 \, , & \lambda N_b \ll 1 \\
\\
\lambda N_b \, , & \lambda N_b \gg 1
\end{array}
\right. \, .
\end{equation}
We compare Eq.~\eqref{eq:DaoudJoanny-Rg2} and Eq.~\eqref{eq:DaoudJoanny-n3} to corresponding results for computer-generated trees in Sec.~\ref{sec:Benchmarking}.

\section{Monte Carlo methods}\label{sec:Methods}
We have employed three different numerical recipes to generate on-lattice random trees subject to the Hamiltonian~\eqref{eq:Hamiltonian} that, essentially, constitute variants of the Amoeba algorithm by Seitz and Klein~\cite{SeitzKlein1981}.
The present Section provides detailed descriptions of these three variants:
\begin{enumerate}
\item
The {\it original} Amoeba algorithm (Sec.~\ref{sec:ConventionalAmoeba}) employs trial moves, where a randomly chosen end node (or leaf) is cut from the tree and reattached to a randomly chosen node of the truncated tree.
\item
The {\it semi-kinetic} Amoeba algorithm (Sec.~\ref{sec:SemiKinAmoeba}) employs trial moves, where again in the first step a randomly chosen leaf is cut from the tree.
In the second step the probability for reattaching it to a node depends on the statistical weight of the resulting tree and hence on whether a branch-point is created (the functionality of the selected node changes from $f\!=\!2$ to $f\!=\!3$) or not (the functionality of the selected node changes from $f\!=\!1$ to $f\!=\!2$).
\item
The {\it double-leaf} Amoeba algorithm (Sec.~\ref{sec:IntroduceDL}) employs additional trial moves affecting peripheral branch-points connecting two leaves to the rest of the tree.
After removing both leaves from a randomly selected branch-point of this type (converting it into a leaf), they are attached to a randomly chosen leaf (converting it into a peripheral branch-point).
\end{enumerate}
The numerical recipes are all based on the Markov Chain Monte Carlo (MC) algorithm, performing sampling of the equilibrium ensemble at constant chemical potential $\mu$ by using the Metropolis-Hastings scheme~\cite{metropolis}.
By enforcing the condition of detailed balance, for any move in the algorithm from an initial configuration $| i \rangle$ to a final configuration $| f \rangle$ the general acceptance probability is given by:
\begin{equation}\label{acc1}
{\rm acc}_{| i \rangle \rightarrow | f \rangle} = \min \left\{ 1, \, \, \, \frac{w(| f \rangle) \, \, \alpha( | f \rangle \rightarrow | i \rangle )}{w( | i \rangle) \, \, \alpha( | i \rangle \rightarrow | f \rangle )}\right \} \, ,
\end{equation}
where $w(| a \rangle)$ is the statistical weight of the state $| a \rangle$ (see Eq.~\eqref{eq:MeanValueGenericObs}).
The $\alpha( | a \rangle \rightarrow | b \rangle)$-terms are algorithm-specific ``attempt probabilities'' for proposing a move from state $| a \rangle$ to $| b \rangle$.
The difference between the algorithms discussed here lies in different definitions  of these attempt probabilities. 

Readers preferring to skip the details of the algorithms on a first reading may advance to Sec.~\ref{sec:Results} and Sec.~\ref{sec:Discussion} for their validation and a discussion of their efficiency for simulating random trees with controlled branching activity.

\subsection{Original Amoeba algorithm}\label{sec:ConventionalAmoeba}
The original version of the Amoeba algorithm is defined by the following steps~\cite{SeitzKlein1981,Rosa2016a}:
\begin{enumerate}
\item
Pick a node A with $f({\rm A})=1$. Node B is the node to which A was attached.
\item
Update the functionality of B as: $f({\rm B})=f({\rm B})-1$.
\item
Pick a node C among all nodes in the chain for which $f({\rm C})<3$. Note: ${\rm C}={\rm B}$ is thus among the possible moves.
\item
Pick an orientation $\vec \gamma$ on the lattice with respect to node C.
\item
Pick a random number $p$. 
\item
If $p < {\rm acc}_{| i \rangle \rightarrow | f \rangle}$, accept the move.
The functionality of node C is updated as $f({\rm C}) = f({\rm C})+1$.
The connectivity of the tree ${\mathcal G}$ is updated by removing the bond between node A and node B and by adding a bond between nodes A and C.
The new coordinate of node A is found by combining the coordinate of node C with the respective orientation $\vec \gamma$.
\end{enumerate}
The acceptance probability (see Eq.~\eqref{acc1}) of the Amoeba algorithm is given by~\cite{SeitzKlein1981}:
\begin{equation}\label{acc2}
{\rm acc}_{| i \rangle \rightarrow | f \rangle} = \min \left\{ 1, \, \, \, \frac{n_1( | f \rangle)}{n_1( | i \rangle )} \, e^{\beta \mu ( n_3 ( | f \rangle) - n_3 ( | i \rangle ) )} \right\} \, .
\end{equation}

In order to efficiently employ the Amoeba algorithm, we introduce the following data structures:
\begin{itemize}
\item
A representation of the connectivity between the nodes ${\mathcal G}$.
In our case, this was an array of directly connected neighbours for every node.
\item
A list of all ``$f\!=\!1$''-nodes, or the nodes to be cut.
\item
A list of all ``$f\!=\!2$''-nodes (for attaching the node to).
\item
A list of all functionalities of the individual nodes.
\item
A list of the coordinates of the individual nodes.
\item
Three integer values $n_1$, $n_2$ and  $n_3$ to keep track of the different type of nodes in the chain.
\end{itemize} 
Since the above algorithm always includes a Markov Chain from any possible tree of $N$ nodes to any other possible tree of $N$ nodes anywhere on the lattice, the algorithm is known~\cite{SeitzKlein1981} to be ergodic (notice, in fact, that the same argumentation holds for the other two algorithms considered in this paper). 

\subsection{Semi-kinetic Amoeba algorithm}\label{sec:SemiKinAmoeba}
Based on Eq.~\eqref{acc2}, it should be clear that in the original Amoeba algorithm some move cases get higher acceptance rates, depending on the sign and absolute value of the chemical potential $\mu$.
Here, we wish to improve this ``asymmetry'' between acceptance probabilities by making the algorithm to propose moves with lower acceptance rates more often as to cancel their statistical weight in the acceptance probability.
This is a method akin to that of the kinetic or ``rejection-free'' MC algorithms which were developed first under different names in the 60's and 70's~\cite{Young1966,Huitson1967,Bortz1975,Gillespie1976}. 

To this purpose, since the MC moves are comprised of two steps, it is possible to define a {\it transition} tree conformation or state $| t \rangle$ as the tree of $N-1$ monomers obtained after cutting a monomer.
With the transition state in hand, the acceptance probability in Eq.~\eqref{acc1} can be factorized as:
\begin{equation}\label{acc3}
{\rm acc}_{| i \rangle \rightarrow | f \rangle} = \min \left\{ 1, \, \frac{w(| f \rangle) \, \alpha( | f \rangle \rightarrow | t \rangle ) \, \alpha( | t \rangle \rightarrow | i \rangle)}{w(| i \rangle) \, \alpha( | i \rangle \rightarrow | t \rangle ) \, \alpha( | t \rangle \rightarrow | f \rangle )} \right \} \, .
\end{equation}
Given a tree in the initial state $| i \rangle$, it is equally probable to cut off any end-point, and similarly for a tree in state $| f \rangle$, therefore:
\begin{equation}\label{eq:probA-ClassicalAmoeba}
\alpha(| i \rangle \rightarrow | t \rangle) = \frac1{n_1(| i \rangle)} \, , \, \, \, \alpha(| f \rangle \rightarrow | t \rangle) = \frac1{n_1(| f \rangle)} \, .
\end{equation}
For any Amoeba MC move, $| i \rangle$ and $| f \rangle$ share the {\it same} transition state $| t \rangle$. 
In particular, the original Amoeba algorithm (Sec.~\ref{sec:ConventionalAmoeba}) has $\alpha(| t \rangle \rightarrow | i \rangle) = \alpha(| t \rangle \rightarrow | f \rangle)$.
Therefore, it is equally likely to pick at random any $1$- or $2$-functional node in $| t \rangle$ to attach a 1-functional node to, before computing the acceptance probability of a move.

Here we propose to change the attempt probabilities~\eqref{eq:probA-ClassicalAmoeba} into the following:
\begin{equation}\label{probA}
\alpha( | t \rangle \rightarrow | i \rangle) = \frac{w( | i \rangle)}{Z_{|t\rangle}} \, , \, \, \, \alpha( | t \rangle \rightarrow | f \rangle) = \frac{w( | f \rangle)}{Z_{|t\rangle}} \, ,
\end{equation}
with $w(\left|i\right\rangle)$ and $w(\left|f\right\rangle)$ being the statistical weights of the original and new configurations (see Eq.~\eqref{eq:MeanValueGenericObs}), respectively, and $Z_{|t\rangle}$ is the partition sum of all states that can be reached from $|t\rangle$.
In this way, the statistical weights drop out in the acceptance probability~\eqref{acc3} that reduces to:
\begin{equation}\label{acc4}
\text{acc}_{| i \rangle \rightarrow | f \rangle} = \min \left\{ 1, \frac{n_1(| i \rangle)}{n_1(| f \rangle)} \right \} \, .
\end{equation}
By comparing the two acceptance probabilities Eq.~\eqref{acc2} and Eq.~\eqref{acc4}, one can expect that, for nonzero $\mu$, the algorithm featuring the latter should benefit from higher acceptance rates.

In order to pick correctly a node C in the transition state, one needs to compute $Z_{|t\rangle}$ (Eq.~\eqref{probA}).
Based on the discussion so far, it is clear that:
\begin{equation}\label{ZT}
Z_{|t\rangle}= n_1(|t\rangle) + n_2(|t\rangle) \, e^{\beta \mu} \, .
\end{equation}
Therefore, depending on the sign of $\mu$, moves that cause branching from the transition state (attachment to a $f=2$ node) will be proposed more or less frequently.
In particular, when branching is energetically penalized ({\it i.e.}, when $\mu<0$), the new algorithm is more likely to choose one of the few end-points to attach a node to.
If it rarely proposes a move that creates a branch, it has a significant chance to be accepted by Eq.~\eqref{acc4}.

In summary, the algorithm reads as the following:
\begin{enumerate}
\item
Pick a node A with $f({\rm A})=1$.
Node B is the node to which A was attached.
\item
Update the functionality of B as: $f({\rm B}) = f({\rm B}) - 1$. 
\item
By using Eq.~\eqref{ZT}, compute $Z_{|t\rangle}$ for the newly obtained transition state.
\item
Pick a random number $q$ between $0$ and $Z_{|t\rangle}$.
\item
If $q < n_1$, choose a node C to attach to between all $1$-functional nodes in the transition state. Otherwise, choose a node C between all $2$-functional nodes. 
\item
Pick an orientation $\vec \gamma$ on the lattice with respect to C.
\item
Pick a random number $p$. 
\item
If $p < {\rm acc}_{| i \rangle \rightarrow | f \rangle}$, accept the move.
The functionality of node C is updated as $f({\rm C}) = f({\rm C})+1$.
The connectivity of the tree ${\mathcal G}$ is updated by removing the bond between node A and node B and by adding a bond between nodes A and C.
The new coordinate of node A is found by combining the coordinate of node C with the respective orientation $\vec \gamma$.
\end{enumerate}
The algorithm presented here needs no additional data structure compared to the original Amoeba algorithm (Sec.~\ref{sec:ConventionalAmoeba}).
The additional steps only require a few extra lines of code that do not slow down the MC steps significantly.
Since we allow the transition state $| t \rangle$ to be chosen at random, without any prior weighting of the leaves to be cut, we named this algorithm the {\it semi-kinetic} Amoeba algorithm, rather than the {\it kinetic} Amoeba algorithm.
The ``fully'' kinetic algorithm ({\it e.g.}, weighting all possible moves of the algorithm on a given polymer) as well as algorithms with different definitions of the transition state $|t \rangle$ are discussed briefly in Sec.~\ref{sec:GeneralizedTransitionStates}.

\subsection{Double-leaf Amoeba algorithm}\label{sec:IntroduceDL}
In this Section we propose to further generalize the Amoeba algorithm. 
The main feature of the new algorithm consists in the ability to move {\it at once} two $1$-functional nodes that are attached to the same node to a different $1$-functional node. 
Hence, this algorithm uses an extended definition of ``leaf", which includes the newly defined ``double leaf''.
In the end, this specific move turns a $1$-functional node into a $3$-functional node (and viceversa).
For the model under investigation, this move does not change the energy of a given chain.
Therefore we have a transition between two chains that have the same statistical weight, {\it i.e.} $w(| i \rangle) = w(| f \rangle)$.
It should be emphasized, though, that, in order to preserve the ergodicity of the algorithm, single-leaf moves need to be preserved.
As such, the double-leaf move is just an additional move.

We formulate the algorithm as the following:
\begin{enumerate}
\item
Pick at random a leaf or double leaf A, among the $n_1$ single leaves with $f\!=\!1$ and the $n_{\rm DL}$ double leaves consisting of two nodes with $f\!=\!1$ attached to the same node with $f\!=\!3$. 
\item
If A is a single leaf, proceed with steps 2-6 in the recipe of Sec.~\ref{sec:SemiKinAmoeba} and then proceed in this recipe from step 7 onwards.
Otherwise, follow steps 3 to 6 described hereunder.
\item
B is the node to which the two nodes of A were attached.
Update the functionality of B as $f({\rm B}) = f({\rm B}) - 2$.
\item
Include B in the list of all $1$-functional nodes of the tree and exclude the $1$-functional nodes in the double leaf to define a transition state $| t \rangle$.
\item
Pick a node C among all $1$-functional nodes in this transition state.
\item
Pick 2 orientations ${\vec \gamma}_1$ and ${\vec \gamma}_2$ on the lattice with respect to C for each node in the double leaf A.
\item
Pick a random number $p$.
\item
If $p < {\rm acc}_{| i \rangle \rightarrow | f \rangle}$, accept the move.
The functionality of node C is updated as $f({\rm C}) = f({\rm C}) + 1$ for a single leaf or as $f({\rm C}) = f({\rm C}) + 2$ for a double leaf.
The connectivity of the tree ${\mathcal G}$ is updated by removing the bond(s) between A and B, and by adding the bond(s) between A and C.
The new coordinate of A is found by combining the coordinate of C with the respective orientation $\vec \gamma$ for a single leaf or with ${\vec \gamma}_1$ and ${\vec \gamma}_2$ for a double leaf.
\end{enumerate}
The recipe described above needs additional data structures compared to the previously described algorithms.
In this work, we have used a list of double leaves and the number of double leaves in a tree ($n_{\rm DL}$).
Besides that, we have used an additional list that keeps track of the number of $1$-functional nodes that were attached to each node.
The acceptance probability can be then computed as:
\begin{equation}\label{acc5}
{\rm acc}_{| i \rangle \rightarrow | f \rangle} = \min\left\{ 1, \, \frac{n_1(| i \rangle) + n_{\rm DL}(| i \rangle)}{n_1(| f \rangle) + n_{\rm DL}(| f \rangle)} \right \} \, .
\end{equation}
Because of the additional technicalities in the data structure, the algorithm presented here is slightly more involved than the algorithms presented before.
One is required to look further into the tree than just the nodes A, B and C, {\it i.e.} the ``active sites" are bigger.
However, when implemented correctly, a MC step will not be slowed down by a factor more than ${\mathcal O}(1)$ compared to the previous algorithms.
Importantly, the extra complication does not cause a $N$-dependent slow-down of the single MC step.

\subsection{Equilibration times}\label{sec:Equilibration}

\subsubsection{Definition}\label{sec:tau_eq_Equilibration}
By construction, leaf-mover algorithms equilibrate trees by cutting and regrowing every bond multiple times in different positions. One can thus monitor the equilibration of the internal structure by following the mean-square displacement for the spatial position, ${\vec R}_{\rm cm} \equiv \frac1N \sum_i^N {\vec r}_i$, of the tree centre of mass~\cite{Rosa2016a,Rosa2016b} as a function of the MC time (also in [MC step] units) $t$,
\begin{equation}\label{eq:Define-g3}
g_3(t) \equiv \left\langle \left( {\vec R}_{\rm cm}(t) - {\vec R}_{\rm cm}(0) \right)^2 \right\rangle  \, .
\end{equation}
As demonstrated in Fig.~1 of Ref.~\cite{Rosa2016a}, equilibration occurs on the time scale where
\begin{equation}\label{g3Rg}
g_3(\tau_{\rm eq}) \simeq \langle R_g^2 \rangle \, .
\end{equation}
Beyond this time
\begin{equation}\label{eq:g3Rg2Scaling}
g_3(t \gtrsim \tau_{\rm eq}) \simeq \langle R_g^2 \rangle \frac{t}{\tau_{\rm eq}} \, ,
\end{equation}
suggesting to use 
\begin{equation}\label{taumax}
t_{\rm eq}\equiv \frac{\langle R_g^2 \rangle}{g_3(t)} \, t 
\end{equation}
as {\it an estimator} of the equilibration time $\tau_{\rm eq}$~\cite{EqTimeInMCsteps}: more precisely, for each $N$ and $\mu$ we compute Eq.~\eqref{taumax} at the end of each single MC trajectory and we average over the independently performed runs. 

\subsubsection{Expected scaling}\label{sec:EquilibrationTimeScaling}
Having introduced the relationship between equilibration and diffusion, it is possible to explore the scaling of the equilibration times in a typical leaf-mover algorithm.
When a leaf is cut from node $i$ of the tree and reattached to node $j$, the tree center of mass moves by a fraction $\delta {\vec R}_{\rm cm} = \frac1N \delta{\vec r}_{ij}$ of the distance between nodes $i$ and $j$. 
Omitting functionality restrictions, the latter is of the order of the tree radius of gyration, 
\begin{equation}
\left. ( \delta{\vec r} )^2 \right|_{\rm single \, move} \simeq \frac1N \frac1{N-1} \sum_i^N \sum_{j \neq i}^N (\delta {\vec r}_{ij})^2 \simeq \langle R_g^2 \rangle \, ,
\end{equation}
so that
\begin{equation}\label{eq:Singlemove}
\left. g_3 \right|_{\rm single \, move} \simeq \frac{\langle R_g^2\rangle}{N^2} \, .
\end{equation}
The combination of Eqs.~\eqref{g3Rg} and~\eqref{eq:Singlemove} suggests the scaling $\tau_{\rm eq}(N) \sim {\mathcal O}(N^2)$ for the equilibration time.
Notice that the result is independent of the internal tree structure and identical to the one expected for linear chain~\cite{deGennes71}.

\section{Results}\label{sec:Results}
The following results were obtained from simulations with the three algorithms presented in Sec.~\ref{sec:Methods} (the original Amoeba algorithm, the semi-kinetic Amoeba algorithm and the double-leaf Amoeba algorithm) of randomly branching trees with $N$ monomers in ideal conditions on the FCC lattice.
In all simulations, the starting configuration was a linear chain. 
We generated $500$ independent tree conformations for given $N \in [ 10,20,40,80,81,161,321 ]$ and $\mu / k_B T \in [ -15, +15 ]$, and each single conformation was obtained by performing a MC run of $1.2 \cdot 10^8$ MC steps that ensures proper equilibration (see Sec.~\ref{sec:Equilibration}) for all systems and the three algorithms under consideration.
For the specific values $N = [ 6,11,21,46,76,151 ]$ and $\mu / k_B T=2$ we have constructed large ensembles of $25000$ independent chains (for the purpose of comparison with the results of Ref.~\cite{Rosa2016b}).
The initial equilibration stage of the trees was removed from the trajectories, to arrive at the results described hereunder. 

\subsection{Validation of the algorithms}\label{sec:Benchmarking}

%
\begin{table*}[htbp]
\begin{tabular}{cccccc}
\multicolumn{6}{c}{$\langle R_g^2 \rangle$} \\
\hline
\hline
\, $N$ \, & \, (2D) \cite{Rosa2016b} \, & \, (3D) \cite{Rosa2016b} \, & \, Original Amoeba \, & \, Semi-kin. Amoeba \, & \, Double-leaf Amoeba \, \\
\hline
6 & $0.845 \pm 0.003$ & $0.851 \pm 0.010$ & $0.844 \pm 0.002$ & $0.849 \pm 0.002$ & $0.846 \pm 0.002$ \\
11 & $1.389 \pm 0.005$ & $1.378 \pm 0.008^\ast$ & $1.386 \pm 0.003$ & $1.381 \pm 0.003$ & $1.388 \pm 0.003$ \\
21 & $2.195 \pm 0.008$ & $2.199 \pm 0.005$ & $2.210 \pm 0.005$ & $2.187 \pm 0.005$ & $2.204\pm 0.005$ \\
46 & $3.684 \pm 0.009$ & $3.688 \pm 0.008$ & $3.682 \pm 0.008$ & $3.670 \pm 0.008$ & $3.676 \pm 0.008$ \\
76 & $5.031 \pm 0.011$ & $5.006 \pm 0.012$ & $5.017 \pm 0.011$ & $4.993 \pm 0.011$ & $4.998 \pm 0.011$ \\
151 & $7.500 \pm 0.022$ & $7.468 \pm 0.016$ & $7.483 \pm 0.017$ & $7.488 \pm 0.017$ & $7.488 \pm 0.017$ \\
\hline
\hline
\\
\multicolumn{6}{c}{$\langle n_3 \rangle$} \\
\hline
\hline
\, $N$ \, & \, (2D) \cite{Rosa2016b}  \, & \,(3D) \cite{Rosa2016b}  \, & \, Original Amoeba \, & \, Semi-kin. Amoeba \, & \, Double-leaf Amoeba \, \\
\hline
6 & $1.429 \pm 0.006$ & $1.450\pm 0.015$ & $1.438 \pm 0.004$ & $1.429 \pm  0.004$ & $1.433 \pm 0.004$ \\
11 & $3.334 \pm 0.006$ & $3.320 \pm 0.012$ & $3.323 \pm 0.004$ & $3.328 \pm  0.004$ & $3.329 \pm 0.004$ \\
21 & $7.292 \pm 0.009$ & $ 7.305 \pm 0.006$ &  $7.293 \pm 0.006$ & $7.290 \pm 0.006$ & $7.292 \pm 0.006$ \\
46 & $17.206 \pm  0.008$ & $ 17.202 \pm 0.008$ &  $17.194 \pm 0.009$ & $ 17.211 \pm 0.009$ & $ 17.211 \pm 0.009$ \\
76 & $29.102 \pm 0.009$ & $29.120 \pm 0.012$ &  $29.099 \pm 0.011$ & $29.106 \pm 0.011$ & $29.103 \pm 0.011$ \\
151 & $58.862 \pm 0.013 $ &$58.874 \pm 0.017$ & $58.872 \pm 0.016$ & $58.864 \pm 0.016$ & $ 58.882 \pm 0.016$ \\
\hline
\hline
\end{tabular}
\caption{
Comparing the three different algorithms by their results for (top) the mean-square gyration radius, $\left\langle R_g^2 \right\rangle$, in (lattice unit = bond length)$^2$ and (bottom) the mean number of branch-points, $\langle n_3\rangle$.
Each row in the tables corresponds to numerical simulations of ideal trees with $N = [ 6, 11, 21, 46, 76, 151 ]$ monomers.
The second and third column of each table show the data appearing in the supplementary material of~\cite{Rosa2016b} obtained on the 2D and 3D simple cubic lattice.
The three columns to the right show the results for the algorithms described in this paper, obtained on the 3D FCC lattice.
The tree observables calculated in this work were obtained by averaging over the independent configurations of $25000$ simulated polymers, and errorbars are estimated from the standard error of the mean.
$^\ast$The value for $\langle R_g^2\rangle$ with $N=11$ in 3D originally reported in~\cite{Rosa2016b} by two of us (AR and RE) was slightly underestimated due to a round-off error that went unnoticed in the course of the automated data analysis. Here we have corrected the mistake, and the new reported value is in agreement with the rest of the other averages. 
}
\label{Rgtable+n3table}
\end{table*}

All three algorithms described in this paper sample the correct ensembles of randomly branching polymers in ideal conditions.

This can be seen first by comparing our results for $\left\langle R_g^2 \right\rangle$ and $\left\langle n_3 \right\rangle$ at $\mu / k_BT = 2$ with those in~\cite{Rosa2016b}, see Table~\ref{Rgtable+n3table}.
Notice that these results refer to 3 different underlying lattices, namely the 2D and 3D simple cubic lattice of Ref.~\cite{Rosa2016b} and the 3D FCC lattice of the present work.
As anticipated in Sec.~\ref{sec:Model}, and as long as trees are studied in the ideal regime, the mean-value of the observables do not depend on the underlying lattice and this is confirmed by the excellent agreement reported in Table~\ref{Rgtable+n3table}.

\begin{figure}[htbp]
\includegraphics[width=0.49\textwidth]{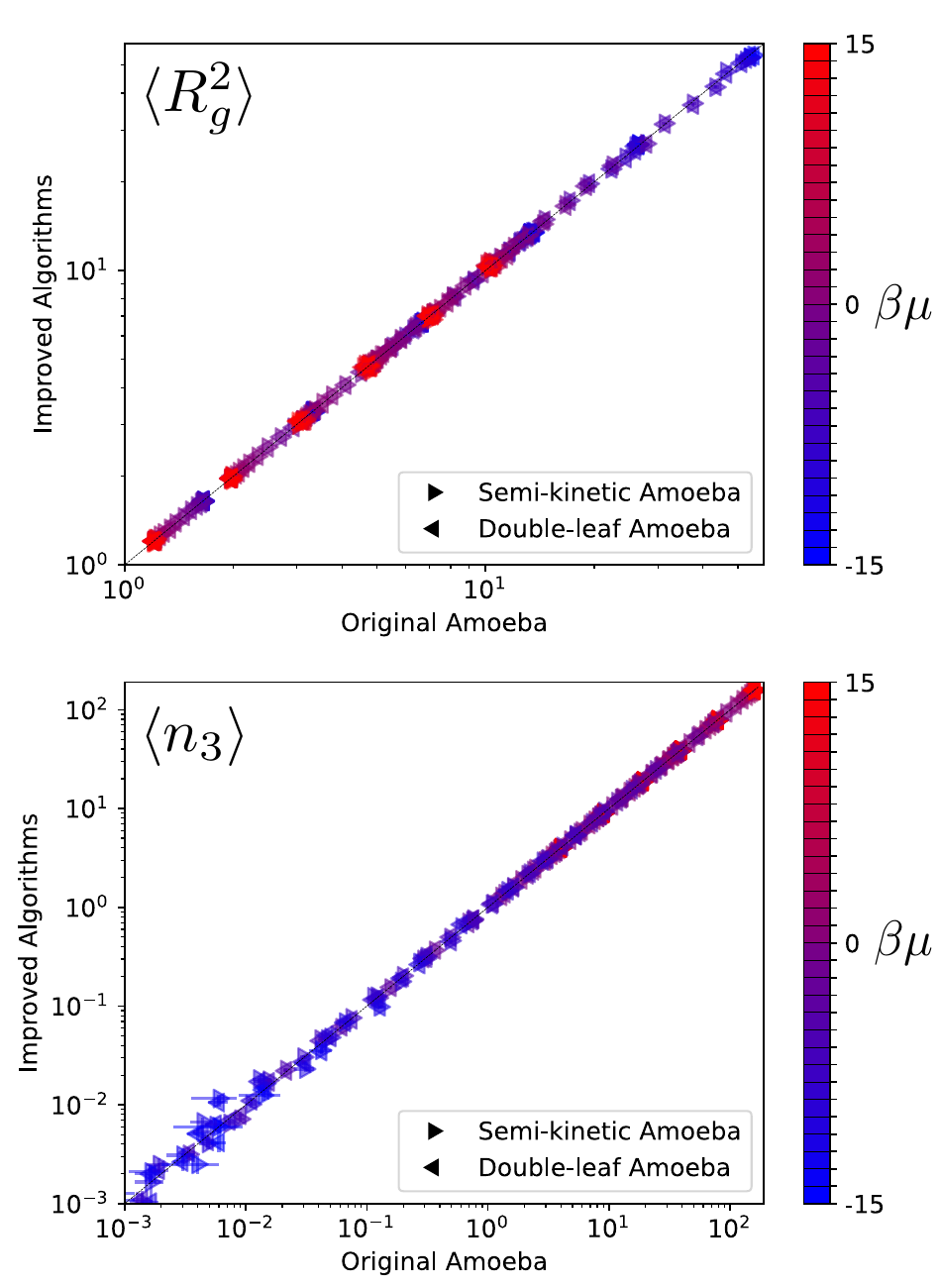}
\caption{
Comparing structural properties of $N$-monomer randomly branching trees for the different algorithms and different values of the branching chemical potential $\mu$ (in units of $\beta^{-1}=k_BT$, see color code bar on the right).
(Top) mean-square gyration radius, $\langle R_g^2\rangle$.
(Bottom) mean number of ``$f\!=\!3$''-nodes, $\langle n_3\rangle$.
The solid line appearing in the background of both panels corresponds to $y=x$.
Error bars are typically smaller than the symbol size.
}
\label{fig:Check_Rg_and_n3}
\end{figure}

Second, we have directly compared independent results for $\langle R_g^2\rangle$ and $\langle n_3\rangle$ from the original and the two new Amoeba algorithms:
as shown in Fig.~\ref{fig:Check_Rg_and_n3}, when plotted against each other results coincide perfectly with the $y=x$ solid line in the background.

\begin{figure}[htbp]
\includegraphics[width=0.49\textwidth]{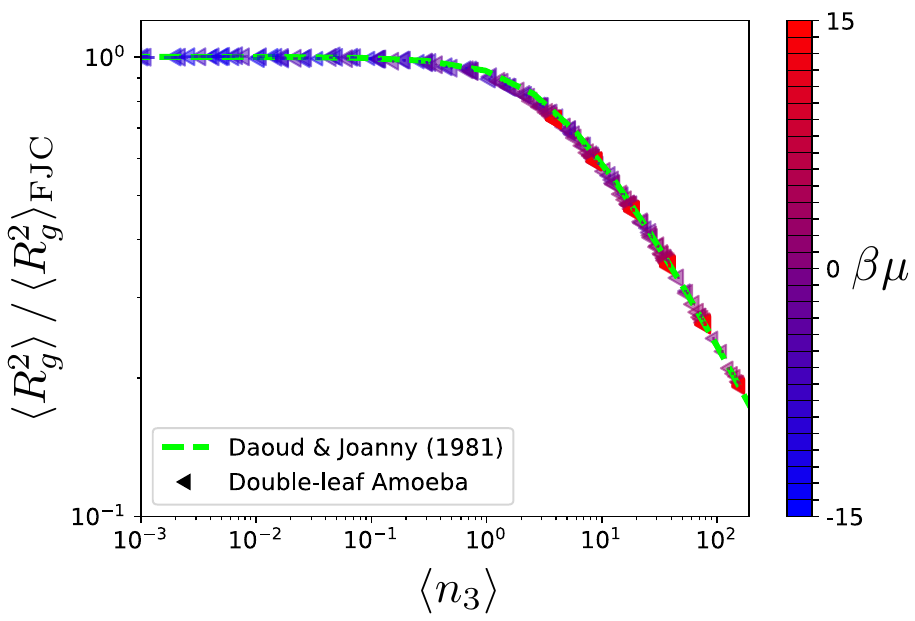}
\caption{
Parametric plot displaying the mean-square gyration radius $\langle R_g^2\rangle$ of $N$-nodes trees normalized to the corresponding value for a freely-jointed chain (FJC) as a function of the total number of branch-points $\langle n_3\rangle$.
Simulation results (symbols) in comparison with the theory (dashed line) for ideal trees by Daoud and Joanny~\cite{DaoudJoanny1981}, see also Sec.~\ref{sec:DaoudJoanny}.
Color code is as in Fig.~\ref{fig:Check_Rg_and_n3}.
Error bars are typically smaller than the symbol size.
}
\label{fig:Daoud_Joanny}
\end{figure}

Third, we have compared results for $\langle R_g^2\rangle$ and $\langle n_3\rangle$ from simulations with the double-leaf Amoeba algorithm with the theoretical predictions of Daoud and Joanny~\cite{DaoudJoanny1981} (see also Sec.~\ref{sec:DaoudJoanny}).
This is shown in Fig.~\ref{fig:Daoud_Joanny}, where $\langle R_g^2\rangle$ (Eq.~\eqref{eq:DaoudJoanny-Rg2}, normalized to the exact value for a freely-jointed chain (FJC, $\lambda N_b \rightarrow0$)) is plotted against the corresponding $\langle n_3\rangle$ (Eq.~\eqref{eq:DaoudJoanny-n3}).
Again, our data (symbols) show perfect agreement with predictions from the continuum theory (dashed line).

\subsection{Equilibration times}\label{sec:EquilibrationTimes}

%
\begin{figure*}[htbp]
\includegraphics[width=1.0\textwidth]{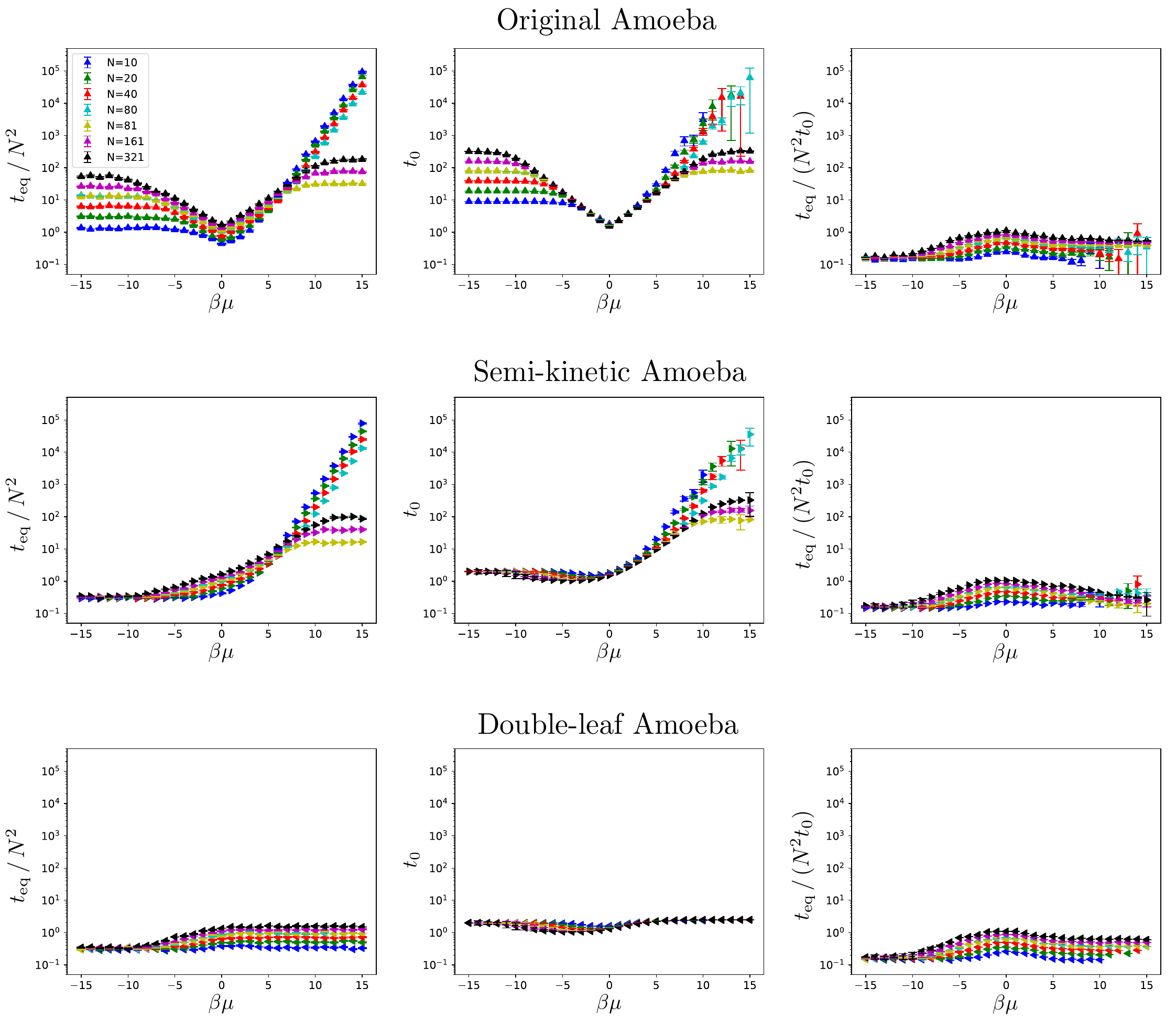}
\caption{
(Left column)
Equilibration times, $t_{\rm eq}$, for trees (according to Eq.~\eqref{taumax}) for the three algorithms of this work (Secs.~\ref{sec:ConventionalAmoeba}-\ref{sec:IntroduceDL}) normalized by $N^2$ as a function of the branching chemical potential $\mu$.
(Middle column)
Estimated time $t_0$ for an elementary connectivity change (Eq.~\eqref{efficientmove} and Table~\ref{tab:move_table}).
Large error bars for some high values of $\mu$ are due to the limited sizes of the corresponding samples.
Datapoints for which no error bars could be defined were left out in these plots. 
(Right column)
Corresponding ratios $t_{\rm eq} \, / (N^2 t_0 )$. 
}
\label{fig:eqreal+eqpred}
\end{figure*}

We are mainly interested in the equilibration or decorrelation times for our simulations (Eq.~\eqref{taumax}).
In the l.h.s. column of Fig.~\ref{fig:eqreal+eqpred} we show our results for $t_{\rm eq} / N^2$ as a function of $\mu$ for the different variants of the Amoeba algorithm. 
The color of the data points serves to distinguish results for different values of $N$.

The original Amoeba algorithm is most efficient for $\mu=0$, when branching is neither promoted nor penalised. 
For odd values of $N$, the observed equilibration times saturate at higher values in both limits of $\mu\ll 0$ (where all trees are just linear chains) and $\mu \gg 0$ (where they are maximally branched). The larger $N$, the higher the values of these plateaus relative to the relaxation times for $\mu=0$.
For $\mu\le 0$ there is no difference in behavior between trees with odd and even numbers of nodes $N$. However, for $\mu>0$ the relaxation times increase exponentially for trees where $N$ is even. 

Are our Amoeba variants more efficient? 
The semi-kinetic Amoeba algorithm is considerably faster for weakly branching trees and recovers the efficiency of the reptation algorithm for linear chains at $\mu\ll0$.
However, it offers essentially no advantage for the investigation of highly branching trees with $\mu\geq 0$.
The remaining difficulty in this limit is overcome by the introduction of double-leaf moves.
The (semi-kinetic) double-leaf Amoeba algorithm fulfils the function for which it was designed: in contrast to the original Amoeba algorithm, it provides an efficient method for generating properly equilibrated samples of branched polymers of arbitrary branching activity, from the weakly- to the highly-branched regime.

\section{Discussion}\label{sec:Discussion}
In this part of the paper we discuss several general aspects of the algorithms studied in this work.
In Sec.~\ref{sec:RelativePerformance} we compare in more details the performance of the three algorithms, while in Sec.~\ref{sec:MovesEfficiency} we look at the role of the different MC moves in leading chains to equilibrium. 
Then, in Sec.~\ref{sec:Universal_dynamics} we reveal a universal tree regime in the Amoeba dynamics that we rationalize by means of scaling arguments.
Finally, in Sec.~\ref{sec:Generalizations} we discuss possible generalizations of the present approach, that include: (i) applications to branch-mover algorithms, (ii) generalized transition states, (iii) trees with nodes of unrestricted functionality and (iv) the introduction of excluded-volume interactions.

\subsection{Relative performance of the three algorithms}\label{sec:RelativePerformance}
From the results on equilibration times in Sec.~\ref{sec:EquilibrationTimes} (Fig.~\ref{fig:eqreal+eqpred}, l.h.s. panels), it is now clear that the three algorithms are comparable in performance only when $\mu=0$.
Here, in fact, the branching Hamiltonian Eq.~\eqref{eq:Hamiltonian} is $=0$, making the original (Sec.~\ref{sec:ConventionalAmoeba}) and the semi-kinetic Amoeba (Sec.~\ref{sec:SemiKinAmoeba}) equivalent to each other (in particular, the attempt and the acceptance probabilities Eqs.~\eqref{acc2} and~\eqref{acc4} are in fact the same for the two algorithms).
As for the introduction of double-leaf moves (Sec.~\ref{sec:IntroduceDL}), the possibility of moving more mass around at once makes the overall performance just slightly better.
On the other hand, for the two regimes where branching is either suppressed ($\mu<0$) or boosted ($\mu>0$) the performance of the original Amoeba algorithm deteriorates rapidly, and it is instructive to explore in more detail why and how the two semi-kinetic algorithms help in that respect.

\subsubsection{$\mu<0$}\label{sec:TauEq-mu<0}
The semi-kinetic algorithms reach the optimal ${\mathcal O}(N^2)$-scaling of the reptation algorithm for $\mu \ll 0$ (where sampled polymers are predominantly linear chains).
The original Amoeba algorithm is a factor of ${\mathcal O}(N)$ slower.
The reason is that it essentially rejects a fraction $\frac{N-2}{N}$ of the attempted moves via Eq.~\eqref{acc2}, because they would create branch-points.
In contrast, branching moves are suppressed in the required proportion in the essentially rejection-free semi-kinetic algorithms.
Introducing double-leaf moves has little effect on equilibration times for $\mu < 0$ (basically, null for $\mu\ll 0$), because the quasi-linear nature of the polymer in this regime obviously excludes the presence of double leaves.

\subsubsection{$\mu>0$}\label{sec:TauEq-mu>0}
If the semi-kinetic algorithm is so much more efficient than the original Amoeba algorithm for $\mu\ll0$, why is there essentially no difference in the opposite limit of $\mu\gg0$, where branching is encouraged (Fig.~\ref{fig:eqreal+eqpred}, l.h.s. column)? 

Consider first a maximally branched tree without a single ``$f\!=\!2$''-functional node, where both algorithms are equally inefficient in changing the conformation: removing a leaf during the first step of a trial move reduces branching, because it converts the adjacent ``$f\!=\!3$''-node to an ``$f\!=\!2$''-node.
In the original Amoeba algorithm, there are two possibilities for the second step.
With a probability of $1/n_1$, the algorithm attempts to attach the leaf to its original site, restoring the original number of branch points.
The trial move is accepted, but does not change the conformation.
With a probability of $1-1/n_1$, the algorithm attempts to attach the leaf to a different leaf, creating a second ``$f\!=\!2$''-node.
Overall, this trial move thus reduces the number of branch points by one.
As it is rejected with high probability, the conformation does not change.
The rejection-free semi-kinetic Amoeba algorithm is ``smarter'' in that it attaches the leaf with high probability to the only ``$f\!=\!2$''-node in the transition state, thereby again restoring the original conformation in the process.
This is the key difference to the linear chain limit, where there are always {\it two} potential attachment points in the transition state: the detached node moves in $50\%$ of the trial moves and the chain relaxes via the characteristic reptation dynamics.

Now assume that we are dealing with a maximally branched tree with a single ``$f\!=\!2$''-functional node.
In the original Amoeba algorithm, a detached leaf will be attached to this node with a probability of $1/n_1$, thereby displacing the $f\!=\!2$-defect to the node to which it was originally detached.
In the semi-kinetic algorithm the defect mobility is much higher, because the detached leaf is relocated with a probability of $50\%$.
However, moving the defect around is insufficient to regrow the tree. 
A branching node becomes mobile {\it only} if it is converted into a leaf such that it can subsequently be moved by the Amoeba algorithm. 
In the present situation this can happen only to peripheral branch points, if two subsequent accepted trial moves remove first one and then the other attached leaf.
The probability that the second trial move removes the second leaf instead of displacing the $f\!=\!2$-defect is $1/n_1$ and identical in both algorithms.
The probability of $n_1/(2n_3+1) \approx 1/2$ that the defect is located on the periphery is algorithm independent.
Thus in total the probability for a relevant change is $1/(2 n_1) \approx 1/N$ in both algorithms.

The importance of the presence of an ``$f\!=\!2$''-site explains the observed difference of the equilibration times for trees with odd and even numbers of nodes in the limit of $\mu\gg0$. 
For odd $N$, there is always at least one ``$f\!=\!2$''-node.
For even $N$, with a the free energy penalty of $\mu$ for the creation of two ``$f\!=\!2$''-nodes, they become exponentially rare.

From the above considerations, it should be clear why double-leaf moves are so effective: they can convert peripheral branch point into leaves in one step and independently of the presence of ``$f\!=\!2$''-nodes.

\subsection{The efficient generation of effective trial moves}\label{sec:MovesEfficiency}

%
\begin{table*}[htbp]
\begin{ruledtabular}
\begin{tabular}{cccccccccc} 
$\rm Class$ & $f({\rm B})$ & $f({\rm C})$ & $\alpha_{\rm O}$ & $\alpha_{\rm SK}$ & $\alpha_{\rm DL}$ & $\text{acc}_{\rm O}$ & $\text{acc}_{\rm SK}$ & $\text{acc}_{\rm DL}$ & $\sigma$\\ 
\colrule
$\rm I$ & 1 & 1 &$\frac{n_{1,2}}{n_1} \frac{n_1-1}{n_1 + n_2 - 1}$ & $\frac{n_{1,2}}{n_1} \frac{n_1-1}{n_1 + (n_2 -1) e^{\mu}}$ & $\frac{n_{1,2}}{n_1} \frac{n_1-1}{n_1 + (n_2 - 1) e^{\mu}} \frac{n_1}{n_1 + n_{\rm DL}}$ & 1 & 1 & 1 & 1 \\
$\rm II$ & 1 & 2 & $ \frac{n_{1,2}}{n_1} \frac{n_2 - 1}{n_2 + n_1 - 1}$ & $\frac{n_{1,2}}{n_1} \frac{(n_2 - 1) e^{\mu}}{ n_1 + (n_2 - 1) e^{\mu}}$ & $\frac{n_{1,2}}{n_1}\frac{(n_2 - 1) e^{\mu}}{ n_1 + (n_2 - 1) e^{\mu}} \frac{n_1}{n_1 + n_{\rm DL}}$ & $\min\left(1, \frac{n_1}{n_1 + 1} e^{\mu}\right)$ & $\frac{n_1}{n_1 + 1}$ & $\frac{n_1 + n_{\text{DL}}}{ n_1 + 1 + n_{\text{DL}}}$ & 1 \\
$\rm III$ & 2 & 1 & $\frac{n_{1,3}}{n_1} \frac{n_1 - 1}{n_1 + n_2}$ & $\frac{n_{1,3}}{n_1}  \frac{n_1 - 1}{n_1 - 1 + (n_2 + 1)e^{\mu}}$ & $\frac{n_{1,3}}{n_1}  \frac{n_1 - 1}{n_1 - 1 + (n_2 + 1) e^{\mu}}\frac{n_1}{n_1 + n_{\rm DL}}$&$\min\left(1,\frac{n_1}{n_1 - 1} e^{-\mu}\right)$& $1$ & $1$  & 1\\
$\rm IV$ & 2 & 2 &$\frac{n_{1,3}}{n_1} \frac{n_2}{n_1 + n_2}$ & $\frac{n_{1,3}}{n_1} \frac{n_2 e^{\mu}}{n_1 - 1 + (n_2 + 1)e^{\mu}}$ &$\frac{n_{1,3}}{n_1 +n_2} \frac{n_2 e^{\mu}}{n_1 - 1 + (n_2 + 1) e^{\mu}}\frac{n_1}{n_1 + n_{\rm DL}}$  & 1 & 1 & 1 & 0 \\
$\rm V$ & 1 & $1^*$ & $\frac{n_{1,2}}{n_1}  \frac{1}{n_1 + n_2 -1}$ & $\frac{n_{1,2}}{n_1} \frac{1}{n_1 + (n_2 - 1)e^{\mu}}$ & $\frac{n_{1,2}}{n_1} \frac{1}{n_1 + (n_2 - 1) e^{\mu}}\frac{n_1}{n_1 + n_{\rm DL}}$& 1 & 1 & 1 & 0 \\
$\rm VI$ & 2 & $2^*$ & $\frac{n_{1,3}}{n_1} \frac{1}{n_1 + n_2}$ & $\frac{n_{1,3}}{n_1} \frac{e^{\mu}}{n_1 - 1 + (n_2+1) e^{\mu}}$ & $\frac{n_{1,3}}{n_1} \frac{e^{\mu}}{n_1 - 1 + (n_2 + 1) e^{\mu}}\frac{n_1}{n_1 + n_{\rm DL}}$& 1 & 1 & 1 & 0 \\
$\rm VII$ & 1 & 1 &  - & - & $\frac{n_{\rm DL}}{n_1 + n_{\rm DL}} \frac{n_1 - 2}{n_1 -1}$ & - & - & 1 & 2 \\
$\rm VIII$ & 1 & $1^*$ &  - & - & $\frac{n_{\rm DL}}{n_1 + n_{\rm DL}}\frac{1}{n_1 -1}$& - & - & 1 & 0 \\
\end{tabular}
\caption{
Characteristics of different classes of Amoeba MC moves discussed in the main text in the various Amoeba algorithms.
$f({\rm B})$ is the functionality of the node which get pruned and $f({\rm C})$ is the functionality of the node that gains a leaf. Both $f({\rm B})$ and $f({\rm C})$ are measured in the transition state $|t\rangle$ (Sec.~\ref{sec:SemiKinAmoeba}).
An asterisk ($^*$) denotes that ${\rm C} = {\rm B}$.
Columns 4-6 portray the {\it attempt} probabilities $\alpha$ in the Original (O), Semi-Kinetic (SK) and Double-Leaf (DL) Amoeba algorithms. 
Columns 7-9 give the approximated {\it acceptance} probabilities for the three algorithms, based on the number of ``$f\!=\!1$''-nodes $n_1$, and the number of double leaves $n_{\rm DL}$.
The ``move effectiveness'' $\sigma$ for the three algorithms as used for our computation of the effective connectivity change per MC time step in Eq.~\eqref{efficientmove} is in column 10.
For simplicity, we approximate that the number of double leaves does not change between two consecutive polymer configurations. 
}
\label{tab:move_table}
\end{ruledtabular}
\end{table*}

The discussion in Sec.~\ref{sec:RelativePerformance} showed that the configurational properties of trees affect the efficiency of the different Amoeba MC algorithms in different ways.
In the following, we explore this quantitatively.

Our Amoeba MC moves fall into eight different classes $\rm I$ to $\rm VIII$, depending on the transition state functionalities and identities of the nodes adjacent to the considered leaves (Table~\ref{tab:move_table}).
Classes $\rm I$ and $\rm V$ regroup the typical reptation moves, where different linear chain segments exchange length (class $\rm I$) or where the detached node returns to its original attached point (class $\rm V$).
Branch points are created and destroyed in class $\rm II$ and $\rm III$ moves. 
Class $\rm IV$ and $\rm VI$ moves displace a leaf from a branch point to a ``$f\!=\!2$''-node (class $\rm IV$) or return it to its previous attachment point (class $\rm VI$).
Classes $\rm VII$ and $\rm VIII$ regroup the double leaf moves, which either attach the two leaves to a new node or to their original attachment point.
Table~\ref{tab:move_table} lists the attempt and acceptance probabilities for the move classes for the considered algorithms as a function of   
the total number $n_1$ of ``$f\!=\!1$''-nodes, 
the total number $n_2$ of ``$f\!=\!2$''-nodes, 
the total number $n_{1,2}$ of ``$f\!=\!1$"-nodes connected to ``$f\!=\!2$''-nodes, 
the total number $n_{1,3}$ of ``$f\!=\!1$"-nodes  connected to ``$f\!=\!3$''-nodes,
and the number  $n_{\rm DL}$ of double leaves.

In particular, we define for each move class an effective connectivity change, $\sigma$, per accepted MC step~\cite{OnClassesAndEfficiency}:
\begin{itemize}
\item
$\sigma_{\rm V}$ = $\sigma_{\rm VI}$ = $\sigma_{\rm VIII}$ = 0: Trivially, these move cases are not effective, because they do not alter the connectivity of a tree.  
\item
$\sigma_{\rm IV}$ = 0: Following the arguments from Sec.~\ref{sec:RelativePerformance}, moves from class $\sigma_{\rm IV}$ are also ineffective.
While they do move individual leaves, they do not mobilise branch points. 
\item
$\sigma_{\rm I} = \sigma_{\rm II} = \sigma_{\rm III} = 1$: All these moves change leaf identities. 
\item
$\sigma_{\rm VII} = 2$: We assume that double-leaf moves are twice as effective than single-leaf moves.
\end{itemize}

Combining the information on the effectiveness and the probability for the occurrence of moves of the different types, we can estimate the {\it effective connectivity change per MC time step},
\begin{equation}\label{efficientmove}
\left(t_0\right)^{-1} = \left. \sum_{\nu = {\rm I}}^{\rm VIII} \, \alpha_{\nu} \, {\rm acc}_{\nu} \, \sigma_{\nu} \right|_{\begin{subarray}{l} \langle n_1\rangle, \, \langle n_2\rangle, \, \\ \langle n_{1,2}\rangle, \, \langle n_{1,3}\rangle, \, \langle n_{\rm DL}\rangle \end{subarray}} \, ,
\end{equation}
for each of the considered algorithms, where $t_0$ denotes an estimate of the number of MC steps required for an `elementary' change in tree connectivity to occur and where we evaluate the expressions in Table~\ref{tab:move_table} using the average values $\langle n_1\rangle$, $\langle n_2\rangle$, $\langle n_{1,2}\rangle$, $\langle n_{1,3}\rangle$, $\langle n_{\rm DL}\rangle$.

Results for $t_0$ shown in the panels in the central column of Fig.~\ref{fig:eqreal+eqpred} are in excellent agreement with the observed reduced equilibration times shown in the l.h.s. column. 
In particular, for our (semi-kinetic) double-leaf Amoeba algorithm, $t_0$ is of the order of a few MC steps and essentially independent of the branching activity.

\subsection{A universal tree regime in the Amoeba dynamics}\label{sec:Universal_dynamics}

%
\begin{figure}[htbp]
\includegraphics[width=0.49\textwidth]{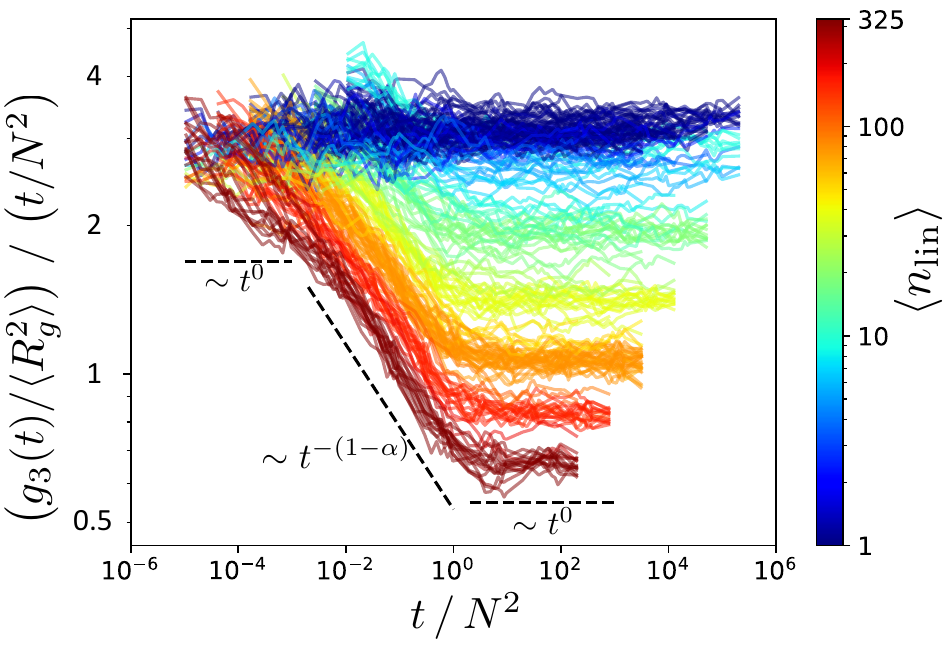} 
\caption{
Trees center-of-mass diffusion, expressed as $\left( g_3(t) / \langle R_g^2 \rangle \right) / \left( t/N^2 \right)$ as a function of $t / N^2$ in the double-leaf Amoeba algorithm.
Color code (see the bar on the right) is according to the mean number of linear segments in the trees, $\langle n_{\rm lin} \rangle = 2 \langle n_3 \rangle +1$.
Error bars are left out for readability.
}
\label{FIG:Scaling_Dynamics}
\end{figure}

The considerations in Sec.~\ref{sec:MovesEfficiency} almost explain the observed equilibration times, but not quite.
This becomes even more clear, when one considers the ratios $t_{\rm eq} \, / \, (N^2 t_0)$ shown in the r.h.s. panels of Fig.~\ref{fig:eqreal+eqpred}.
They suggest an additional, weak, algorithm-independent and hence universal contribution to the $N$-dependence of the Amoeba equilibration times beyond the expected reptation-like ${\mathcal O}(N^2)$ scaling. 
Considering how we have defined the equilibration time in Sec.~\ref{sec:Equilibration}, we need to explore the tree center-of-mass diffusion at earlier times, if we want to better understand deviations from the expected reptation-like scaling Eq.~\eqref{eq:Singlemove}.

To this end, we show in Fig.~\ref{FIG:Scaling_Dynamics} the ratio of the measured and the expected center-of-mass diffusion $g_3(t)$ for all investigated combinations of $N$ and $\mu$ in the double-leaf Amoeba algorithm.
All data appear to collapse at the earliest times.
Most data sets exhibit an early time regime, where $( g_3 / \langle R_g^2\rangle ) \, / \, (t / N^2)$ is independent of time. 
All systems also reach a terminal diffusion regime, where $( g_3 / \langle R_g^2\rangle ) \, / \, (t / N^2)$ is again independent of time.
Interestingly most systems exhibit an intermediate anomalous diffusion regime ($g_3 \sim t^{\alpha}$, $\alpha<1$) and thus diffuse more slowly on long than on short time scales.

Importantly, the color coding reveals that the deviations from ordinary diffusion are entirely controlled by the mean number 
\begin{equation}\label{eq:n_lin}
\langle n_{\rm lin} \rangle = 2 \langle n_3 \rangle + 1
\end{equation}
of linear segments in the trees. 
Linear chains with $\langle n_{\rm lin}\rangle =1$ exhibit the expected time-independent center-of-mass diffusion. 
For branched trees, the larger $\langle n_{\rm lin} \rangle \gg 1$, the longer the anomalous diffusion regime and the larger the slowdown of the asymptotic diffusion.

To rationalise these observations, we introduce the following scaling form for $g_3(t)$ of trees in the double-leaf Amoeba algorithm: 
\begin{equation}\label{eq:g3-scaling}
g_3(t) \simeq \langle R_g^2 \rangle \times
\left\{ \begin{array}{cc} \frac1{N^2} \frac{t}{\tau_0} \, , & t\lesssim \tau_1 \\ \\ \langle n_{\rm lin}\rangle^x \left(\frac{t}{\tau_1}\right)^{\alpha} \, , & \tau_1 \lesssim t \lesssim \tau_{\rm eq} \\ \\ \frac{t}{\tau_{\rm eq}} \, , & t \gtrsim \tau_{\rm eq} \end{array} \right. \, .
\end{equation}
with
\begin{eqnarray}
\tau_1 & = & N^2 \langle n_{\rm lin}\rangle^x \, \tau_0 \, , \label{eq:tau_1} \\
\tau_{\rm eq} & = & N^2 \langle n_{\rm lin}\rangle^{\Delta} \, \tau_0 \, , \label{eq:tau_eq}
\end{eqnarray}
where $\tau_0$ is a microscopic time from the point of view of the tree dynamics.
In Eq.~\eqref{eq:g3-scaling} the intermediate time scale $\tau_1$ separates (i) the early regime during which the algorithm predominantly performs mass exchange between the outer linear segments of the trees from (ii) the tree regime controlled by the destruction and creation of branch-points, during which the tree centers-of-mass exhibit anomalous diffusion.
Notice that the two proposed scaling behaviors for $\tau_1$ and $\tau_{\rm eq}$ in Eqs.~\eqref{eq:tau_1} and~\eqref{eq:tau_eq} follow from the mentioned observation that $\langle n_{\rm lin}\rangle$ controls the deviations of $g_3$ from pure diffusion.
In particular, the two characteristic time scales converge to the standard reptation expression in the absence of branch points: $\lim_{\langle n_{\rm lin} \rangle \rightarrow 1} \tau_1 = \lim_{\langle n_{\rm lin} \rangle \rightarrow 1} \tau_{\rm eq} = N^2 \tau_0$.

\begin{figure*}[htbp]
\includegraphics[width=0.90\textwidth]{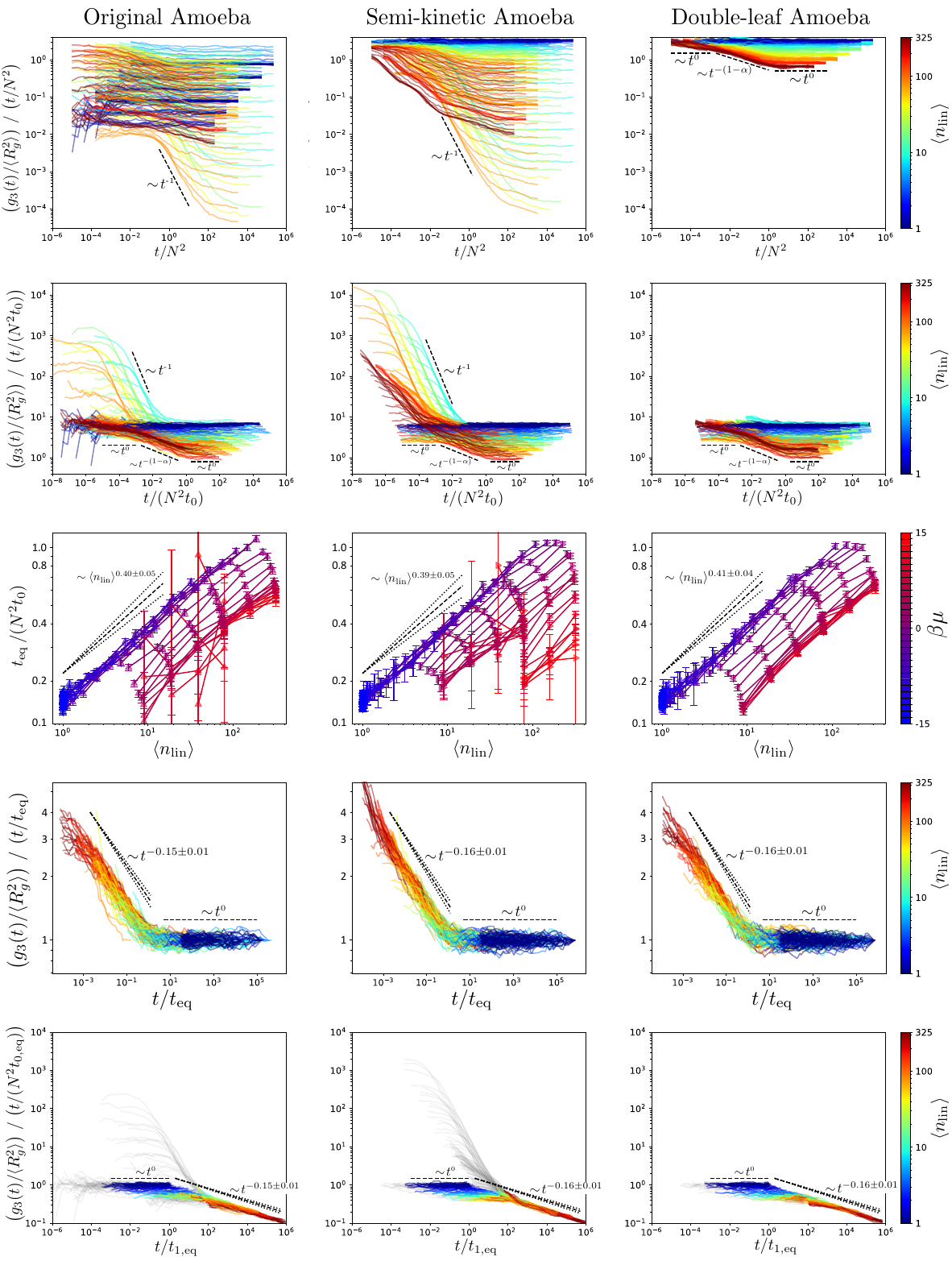}
\caption{
Scaling analysis (see Sec.~\ref{sec:Universal_dynamics} for details) of trees center-of-mass diffusion, $g_3(t)$, in the original (left column), semi-kinetic (center column) and double-leaf (right column) Amoeba algorithms.
Color code in rows (top to bottom) 1, 2, 4 and 5 is as in in Fig.~\ref{FIG:Scaling_Dynamics}, while in row 3 color code is as in Fig.~\ref{fig:Check_Rg_and_n3}.
In all plots, except in row 3, error bars are not displayed in favor of readability.
All plots include proportionality lines to guide the eye.
(Row 1)
``Raw'' diffusion data, expressed as $( g_3(t) / \langle R_g^2 \rangle ) \, / \, ( t / N^2 )$ as a function of $t / N^2$ ({\it i.e.}, as in Fig.~\ref{FIG:Scaling_Dynamics}).
(Row 2)
Same data as in Row 1, with time expressed in units of the ``elementary'' time scale for effective connectivity change ($t_0$, Eq.~\eqref{efficientmove}).
(Row 3)
Equilibration times $t_{\rm eq} \, / (N^2 t_0)$ as a function of the mean number of linear segments $\langle n_{\rm lin} \rangle$ of the trees.
Datapoints for the same $\mu$ are connected by lines to guide the eye. 
As in Fig.~\ref{fig:eqreal+eqpred}, datapoints with poor statistics are excluded from the plots.
(Row 4)
Plots of $( g_3(t) / \langle R_g^2 \rangle ) \, / \, ( t / t_{\rm eq} )$ as a function of $t / t_{\rm eq}$.
Data for $t / t_0 < 5 ( N / \langle n_{\rm lin} \rangle )^2$ (see Eq.~\eqref{eq:tau_1_for_x=2}) are not displayed. 
(Row 5)
Plots of $( g_3(t) / \langle R_g^2 \rangle ) \, / \, ( t / (N^2 t_{0, \rm eq}) )$ as a function of $t / t_{1, \rm eq}$ where $t_{0, \rm eq}$ and $t_{1, \rm eq}$ are defined in Eq.~\eqref{t0_eq} and Eq.~\eqref{t1_eq}, respectively.
Data for $t > t_{\rm eq}$ are not displayed.
The algorithmic-specific lines portions are grey-colored for clarity.
}
\label{FIG:Universal_Dynamics}
\end{figure*}

The exponents $\alpha$, $x$ and $\Delta$ are not independent from each other as the continuity of Eq.~\eqref{eq:g3-scaling} at $t=\tau_{\rm eq}$ imposes that
\begin{equation}\label{eq:g3-scaling-ExpsRelation}
\Delta = -\frac{1-\alpha}\alpha x \, .
\end{equation}
In the following we combine the insight from the scaling ansatz~\eqref{eq:g3-scaling} with our results from the previous Sec.~\ref{sec:MovesEfficiency} to arrive at a common interpretation of the data from the three investigated Amoeba algorithms (represented in Fig.~\ref{FIG:Universal_Dynamics} in separate columns and with common scales for corresponding panels).

On first sight, the raw data for the other algorithms are not very promising when plotted in the same way as Fig.~\ref{FIG:Scaling_Dynamics} (top row of Fig.~\ref{FIG:Universal_Dynamics}), but one can nevertheless recognise a few features.
For all three algorithms results for linear chains (color coded in blue) show up as horizontal lines, indicative of time-independent diffusive behavior.
For the semi-kinetic algorithm, the data collapse as in the case of the double-leaf algorithm for the simple reason that the two are essentially identical for linear chains.
For the original Amoeba algorithm, the blue horizontal lines are shifted relative to each other due to the ${\mathcal O}(N)$ rejected side-branching attempts between successful reptation moves discussed in Sec.~\ref{sec:TauEq-mu<0}. 
Similarly, data for highly branched trees (color coded in orange or red) appear shifted for the original and the semi-kinetic Amoeba algorithms, because of their difficulties in mobilising the branch points. The extreme cases of maximally branched trees with even $N$, where the number of ``$f\!=\!2$''-functional nodes becomes exponentially small, manifest themselves by plateaus in $g_3(t)$ over extended waiting times, which translate to $g_3(t)/t \sim 1/t$ in the present representation. 

The panels in the second row in Fig.~\ref{FIG:Universal_Dynamics} show that the differences between the algorithms largely vanish, when time is measured in units of the specific values of $t_0$ from Eq.~\eqref{efficientmove}.
This shows that
(i) the increases of equilibration times in the original and semi-kinetic Amoeba algorithm compared to the double-leaf Amoeba algorithm are due to short time-scale effects
and that
(ii) the algorithms display a {\it universal} Amoeba dynamics described by the scaling theory in Eq.~\eqref{eq:g3-scaling} when time is measured in units of $t_0$.
In the following we proceed by extracting the characteristic exponents $\Delta$, $\alpha$ and $x$ from the three data sets. 

The panels in the third row of Fig.~\ref{FIG:Universal_Dynamics} show measured equilibration times $t_{\rm eq} \, / (N^2 t_0)$ in units of $t_0$ as a function of $\langle n_{\rm lin}\rangle$.
Lines connect data points for identical values of $\mu$ and are themselves color coded as a function of $\mu$. 
Individual data sets are compatible with our postulated (Eq.~\eqref{eq:tau_eq}) power law dependence of $t_{\rm eq}/N^2 \sim \langle n_{\rm lin}\rangle^\Delta$.
They do not superpose perfectly, because our estimates of $t_0$, which vary over 5 orders of magnitude (central column of Fig.~\ref{fig:eqreal+eqpred}), are only correct within a factor of two or three.
Combining individual fits of the slopes we arrive at estimates of
$\Delta = 0.40 \pm 0.05$ (original Amoeba),
$\Delta = 0.39 \pm 0.05$ (semi-kinetic Amoeba)
and
$\Delta = 0.41 \pm 0.04$ (double-leaf Amoeba) for the three algorithms, which agree within the estimated statistical and systematic errors.

The panels in the forth row of Fig.~\ref{FIG:Universal_Dynamics} show scaling plots of $( g_3(t) / \langle R_g^2\rangle ) \, / \, (t / t_{\rm eq})$, where time is rescaled with the measured equilibration times and where we have excluded data from the early time regime.
This representation focuses on the transition from the subdiffusive to the terminal diffusive regime.
For all three algorithms we observe a perfect data collapse between data for different $\mu$.
Furthermore, the results for the three algorithms are nearly indistinguishable and can be superposed (data not shown).
From Eq.~\eqref{eq:g3-scaling} with Eq.~\eqref{eq:g3-scaling-ExpsRelation} we expect 
\begin{equation}\label{eq:g3-scaling_at_end}
\frac{g_3(t) / \langle R_g^2 \rangle}{t/\tau_{\rm eq}} \simeq \left\{ \begin{array}{cc} \left(\frac{t}{\tau_{\rm eq}}\right)^{-(1-\alpha)} \, , & \tau_1 \lesssim t \lesssim \tau_{\rm eq} \\ \\ 1 \, , & t \gtrsim \tau_{\rm eq} \end{array} \right. \, .
\end{equation}
Again we have extracted estimates of the exponent $\alpha$ from the data and found with $\alpha = 0.85 \pm 0.01$ (original Amoeba), $\alpha = 0.84 \pm 0.01$ (semi-kinetic Amoeba), and $\alpha = 0.84 \pm 0.01$ (double-leaf Amoeba) compatible values for the three algorithms.

Eq.~\eqref{eq:g3-scaling-ExpsRelation} allows to convert the measured values of $\Delta$ and $\alpha$ into a common estimate $x = -2.2 \pm 0.2$ of the third exponent $x$ for Amoeba algorithms.
Interestingly Eq.~\eqref{eq:tau_1} for $x=-2$ reads
\begin{equation}\label{eq:tau_1_for_x=2}
\frac{\tau_1}{\tau_0} = \frac{N^2}{\langle n_{\rm lin}\rangle^2} = N_{\rm lin}^2 \, ,
\end{equation}
{\it i.e.} it is given by the square of the number of nodes per linear segment and is, for a given $\mu$ and up to finite-size corrections, independent of the tree weight. 

With $\tau_1 / \tau_{\rm eq} = \langle n_{\rm lin}\rangle^{x-\Delta}$ (Eqs.~\eqref{eq:tau_1} and~\eqref{eq:tau_eq}) we can use our measured values of $t_{\rm eq}$ to define estimates
\begin{eqnarray}
t_{0, \rm eq} & = & t_{\rm eq} \, / \left( N^2 \langle n_{\rm lin} \rangle^{0.4} \right) \, , \label{t0_eq} \\
t_{1, \rm eq} & = & t_{\rm eq} \, / \, \langle n_{\rm lin} \rangle^{2.4} \, , \label{t1_eq}
\end{eqnarray} 
of the time scale $\tau_0$ and of the time scale $\tau_1$, which controls the onset of the subdiffusive regime.
From Eq.~\eqref{eq:g3-scaling} with Eq.~\eqref{eq:g3-scaling-ExpsRelation} we expect for times around $\tau_1$: 
\begin{equation}\label{eq:g3-scaling_at_onset}
\frac{g_3(t) / \langle R_g^2 \rangle}{t / (N^2 \tau_0)} \simeq \left\{ \begin{array}{cc} 1 \, , & t \lesssim \tau_1 \\ \\ \left(\frac{t}{\tau_1}\right)^{-(1-\alpha)} \, , & \tau_1 \lesssim t \lesssim \tau_{\rm eq} \end{array} \right. \, .
\end{equation}
The panels in the last row of Fig.~\ref{FIG:Universal_Dynamics} show scaling plots of $( g_3(t) / \langle R_g^2\rangle ) \, / \, (t / t_{0, \rm eq})$, which align the data on the crossover into the tree regime.
This time we exclude data {\it beyond} $t_{\rm eq}$, while data for the non-universal early-time dynamics for $t < 5 t_0$ are shown in gray. 
The data collapse is satisfactory, but less convincing than for the end of the tree regime.
For highly branching trees with $N_{\rm lin} = 1$, the early-time dynamics crosses over directly into the subdiffusive regime. 
For trees with long linear segments, $\tau_1 \gg \tau_0$ and we observe the expected extended initial diffusion regime. 

It is worthwhile mentioning that a similar power law correction for the equilibration times of trees relative to the reptation estimate was reported in a study by Janse van Rensburg and Madras~\cite{MadrasJPhysA1992}.
At the time, the authors ascribed the slow-down to excluded volume interactions in the studied self-avoiding trees.
Here, we have shown that this effect also exists for ideal trees and that it is probably a general characteristic of Amoeba algorithms.
We conclude that equilibration times of the functional form~\eqref{eq:tau_eq} ought to be expected in {\it any} Amoeba-like algorithm.

That being said, a quantitive explanation for the observed value $\Delta \simeq 0.4$ remains to be found.
The existence of an intermediate time scale $\tau_1$, from which subdiffusive behaviour sets in, indicates that the growing and pruning of entire branches displays more correlated behaviour than what can be na\"ively assumed from reptation arguments (Sec.~\ref{sec:Equilibration}) alone.
A possible speculation is that, in trees, the nodes that are being moved around by the algorithm at intermediate times ({\it i.e.}, for $\tau_1 < t < \tau_{\rm eq}$) are often ``taken'' from the same subset of all $N$ nodes in the tree.
This effect should then make it harder for the algorithm to reach the tree central node, compared to when reptation is applied to a linear chain. 

\subsection{Possible generalizations}\label{sec:Generalizations}

\subsubsection{Branch-mover algorithms}\label{sec:BranchMovingAlgos}
Inspired by the original MC Amoeba algorithm by Seitz and Klein~\cite{SeitzKlein1981}, the algorithms we discussed here are all {\it leaf}-mover algorithms. 
Yet, these algorithms are not necessarily the fastest ones when it comes to equilibration of single, isolated chains. 
In fact, as originally reported by Janse van Rensburg and Madras~\cite{MadrasJPhysA1992} concerning single trees with self-avoiding interactions, a better strategy consists in employing {\it branch}-mover algorithms where tree branches of arbitrary size (and not just leaves) are cut and resealed elsewhere.
However, Janse van Rensburg and Madras~\cite{MadrasJPhysA1992} did not consider trees with controlled branching activity: after including this term into the model, we expect the original branch-mover algorithm to suffer from the same difficulties as the original leaf-mover algorithm. 
In other words, cutting and attaching branches at random as proposed by Janse van Rensburg and Madras would not prevent frequent rejections in the linear and highly-branched regimes.

Fortunately, generalizing our methods to branch-moving algorithms is straightforward: namely, a transition state as introduced in Sec.~\ref{sec:SemiKinAmoeba} can be reached by pruning a tree from any node at random.
For this transition state, a similar semi-kinetic prescription can be implemented in order to make the algorithm approximately rejection-free as well.
Next, the introduction of double-leaf moves into the algorithm would already avoid any exponential increase in equilibration times for higher positive $\mu$.
In addition, one might even choose to cut {\it double branches} from any $3$-functional node in the tree.

 \subsubsection{Generalized transition states}\label{sec:GeneralizedTransitionStates}
The introduction of a transition state $| t \rangle$ in Sec.~\ref{sec:SemiKinAmoeba} proved to be a useful step in obtaining an algorithm with faster equilibration times.
However, our definition of $| t \rangle$ is not unique.
We implemented variations of this method, but they are in general not as effective as the methods discussed in this paper. We will argue below why this was the case.

As an example, we considered to first attach a bond at random to create a transition state with $N+1$ nodes and then cut off a $1$-functional node, by using a similar function as Eq.~\eqref{ZT}.
This approach is not very effective in the linear regime, as it is very likely to create a branch-point that will be cut off in the same MC move, leaving the chain unchanged.
In the highly branched regime, the method is not very effective for similar reasons: there are more $1$-functional nodes than $2$-functional ones, so that most random attachments would create linear segments with an exponentially higher chance to be removed back as well.
For these reasons, we did not explore such an algorithm any further. 

A possible alternative is to switch from a ``semi-kinetic'' to a ``kinetic'' method: by no longer cutting monomers at random but statistically weighting every possible move that a given tree allows, we can skip the entire transition state all at once.
This method has, however, several disadvantages.
First of all, the number of classes of moves increases considerably.
Instead of choosing between attachment to a $1$-functional or a $2$-functional node, one has to check all possible combinations of cutting and attaching nodes, not only for the initial tree, but also for the final tree.
In total this gives $16$ classes of moves for our model with restricted functionality.
The increase in classes of moves results in an increase of the amount of updates of relevant data structures per MC step.
The latter, in turn, shows up as an increase of CPU-time not present in the other algorithms discussed here.
An additional problem arises because the partition functions as in Eq.~\eqref{ZT} for the initial and the final tree no longer cancel out in the acceptance probability.
Hence an expression for the acceptance probability as Eq.~\eqref{acc4} is no longer achievable, and it instead looks like:
\begin{equation}
\text{acc}_{| i \rangle \rightarrow | f \rangle} = \min \left\{ 1, \frac{Z_{|i\rangle}}{Z_{|f\rangle}} \right \} \, ,
\end{equation}
when moves are weighted such to cancel their statistical weight.
This is particularly detrimental in the maximally branched regime, since any final state $|f \rangle$ holding a linear segment when $|i \rangle$ has none can only be reached with a ${\mathcal O}(e^{-\mu})$ chance.
Moves can be reweighed differently as well, but the non-cancelation of the partition functions in the acceptance probability will remain regardless of the reweighting choice.
For this reason, and for its considerable complexity in comparison to the algorithms presented in this work, we did not investigate such method further.

\subsubsection{Branch-points with higher functionality}\label{sec:NodesUnrestrictedFuncts}
The most general Hamiltonian for controlling the tree composition reads
\begin{equation}\label{eq:Hamiltonian_all_f}
\mathcal{H}({\mathcal T}) = - \sum_{f=1}^{f_{\rm max}} \mu_f^{(0)} \, n_f({\mathcal G}) \, ,
\end{equation}
and allows for the definition of individual biases controlling the presence of endpoints ($f=1$), linear tree segments ($f=2$), and branch-points of particular functionalities $f>3$.
Alternatively a term $\mu_f\equiv \mu_{\rm br} \, , \, \forall f\ge3$ could control the presence of branch-points independently of their functionality. 

For our choice of $f_{\rm max}=3$, Eqs.~\eqref{n3n1} and~\eqref{n3n2} allow to reduce Eq.~\eqref{eq:Hamiltonian_all_f} to Eq.~\eqref{eq:Hamiltonian} with a single parameter $\mu = \mu_1^{(0)} -2\mu_2^{(0)} +\mu_3^{(0)} $ accounting for the combined effect of the individual biases. 
For models authorizing branch-points to have higher functionalities, Eqs.~\eqref{n3n1} and~\eqref{n3n2} generalise to
\begin{eqnarray}
n_1 & = & 2 + \sum^{N-2}_{f>2} (f-2) \, n_f \, \label{n3en1} \, , \\
n_2 & = & N - \sum^{N-2}_{f>2} (f-1) \, n_f - 2 \,  \label{n3en2} \, ,
\end{eqnarray}
so that it is again possible to work in a gauge where the effective chemical potentials for one- and two-functional nodes are set to $\mu_1=\mu_2=0$ and 
\begin{eqnarray}
\mu_f =  (f-2)\mu_1^{(0)}  - (f-1)\mu_2^{(0)}  + \mu_f^{(0)} \, .
\end{eqnarray}
Such models display a rich behavior.
While their study is beyond the scope of the present work, we note that the present algorithmic ideas can be generalised to these case as well.
The semi-kinetic choice of trial moves, Eq.~\eqref{probA}, can be formulated for any set  $\{\mu_f\}$ of (effective) branching potentials and the double-leaf move generalizes to multiple-leaf moves for random numbers of leaves connected to a common branch-point.
Although this might increase the complexity of the data structures, we think the generalizations are straightforward and would not lead to a significant increase in CPU times.

\subsubsection{Excluded volume interactions}\label{sec:EVDiscussion}
The original Amoeba proved particularly efficient in equilibrating interacting trees at dense (melt) conditions and for a polymer model with soft excluded volume interactions~\cite{Rosa2016b}.
We expect that the solutions provided here ought to be able to equilibrate chain connectivity and spatial conformations even more efficiently than the original Amoeba algorithm and even when excluded volume effects are included, particularly in those regimes where we have shown that one can gain a factor ${\mathcal O}(N)$ or a factor ${\mathcal O}(e^{\mu})$ in equilibration times.

Of course, it remains plausible that the improvements presented here might be slightly diminished when including excluded volume effects.
For instance, moves that equilibrate the {\it connectivity} of the chain ${\mathcal G}$ might excite the chain energy through {\it spatial} interactions, and it might be necessary to move through an excited state in the connectivity ${\mathcal G}$ to lower the energy of a given spatial configuration $\Gamma$.
In this respect, the semi-kinetic Amoeba algorithm might therefore not work as efficiently as seen in the ideal situation.
Moreover, when bringing two nodes from a dilute region to a denser region, the double-leaf Amoeba algorithm has a higher risk for rejection than when only one leaf is moved around.

For all these reasons, it is an exciting question to extend the same philosophy of this work to trees with excluded volume as well, particularly in melt conditions.
We believe that this should be possible, and we will present it in future work.

\section{Conclusion and outlook}\label{sec:Conclusion}
In this work, we have generalized the original Amoeba algorithm for random trees~\cite{SeitzKlein1981} with the objective of efficiently generating lattice trees with controlled branching activities of arbitrary strength.
For low branching activity ({\it i.e.}, negative $\mu$ in the branching Hamiltonian Eq.~\eqref{eq:Hamiltonian}) we suggest a rejection-free semi-kinetic algorithm for choosing branching moves with a suitable probability, while for high branching activity ({\it i.e.}, positive $\mu$) we propose to move multiple leaves at the same time in a single MC step. 
The (semi-kinetic) double-leaf Amoeba algorithm fulfils the function for which it was designed: it allows to generate properly equilibrated samples of branched polymers of arbitrary branching activity with optimal efficiency all the way from the weakly- to the highly-branched regime.

A detailed analysis of chain relaxation to equilibrium reveals an unexpectedly rich behavior of the equilibration time, $\tau_{\rm eq}$, as a function of both total number of nodes ($N$) and mean number of branching nodes ($\langle n_3\rangle$), that can not be explained by simple arguments of non-local mass transport across the chain.
Then, by means of simple scaling arguments, we show that the reported behavior for $\tau_{\rm eq}$ is consequent on the subdiffusive behavior of the tree centre of mass that sets in at the characteristic time scale, $\tau_1$, for the early reptation-like regime for the rearrangement of individual linear segments.

In perspective, the numerical solutions provided in this work can be straightforwardly implemented and applied to study a wide range of physical systems.
In particular to study dense melts of trees, an important class of systems that serves as a role model for understanding the spatial behavior of ring polymer melts~\cite{RosaEveraers2019,Ghobadpour2021} and DNA organization in eukaryotes~\cite{RosaPLOS2008,RosaEveraersPRL2014} and where the original Amoeba algorithm has been already applied with proficiency~\cite{Rosa2016b}.

{\it Acknowledgements} --
PHWvdH acknowledges financial support from PNRR\_M4C2I4.1.\_DM351 funded by NextGenerationEU and the kind hospitality of the ENS-Lyon.
AR acknowledges financial support from PNRR Grant CN\_00000013\_CN-HPC, M4C2I1.4, spoke 7, funded by Next Generation EU.

\bibliography{biblio.bib}

\begin{thebibliography}{72}
\bibitem [{EqT()}]{EqTimeInMCsteps}%
  \BibitemOpen
  \href@noop {} {}\bibinfo {note} {We have chosen to measure performance in MC
  steps and not in CPU-time, since the computational efforts for a single MC
  step in the different algorithms are comparable
  (Secs.~\ref{sec:ConventionalAmoeba}-\ref{sec:IntroduceDL}).}\BibitemShut
  {Stop}%
\bibitem [{OnC()}]{OnClassesAndEfficiency}%
  \BibitemOpen
  \href@noop {} {}\bibinfo {note} {Despite being classified as
  ``non-effective'' in this part of the text, all move classes for which
  $\sigma= 0$ are essential for the semi-kinetic procedure described in
  Sec.~\ref{sec:SemiKinAmoeba}. Namely, in order to have the same transition
  state $| t \rangle$ between an initial state $| i \rangle$ and a final state
  $| f \rangle$, all move classes are necessary. Only with the same transition
  state $| t \rangle$, one can reach the efficient acceptance probabilities in
  Eqs.~\eqref{acc4} and~\eqref{acc5}. Moreover, due to the move classes for
  which $\sigma=0$ still being capable of altering spatial positions of
  individual nodes, they are of importance as soon as spatial interactions are
  introduced.}\BibitemShut {Stop}%
\end{thebibliography}

%
\end{document}